\documentclass[%
 rsi, superscriptaddress,
 amsmath,amssymb,
 reprint,%
]{revtex4-1}

\usepackage{graphicx}
\usepackage{dcolumn}
\usepackage{bm}

\usepackage[utf8]{inputenc}
\usepackage[T1]{fontenc}
\usepackage{mathptmx}
\usepackage{hyperref}

\begin{document}

\title{Compact integrated optical sensors and electromagnetic actuators for vibration isolation systems in the gravitational-wave detector KAGRA}

\author{Tomotada Akutsu}
\email{tomo.akutsu@nao.ac.jp}
 \affiliation{National Astronomical Observatory of Japan (NAOJ), Mitaka, Tokyo 181-8588, Japan}
 \affiliation{The Graduate University for Advanced Studies (Sokendai), Mitaka, Tokyo 181-8588, Japan}
\author{Fabián Erasmo Peña Arellano}
 \affiliation{National Astronomical Observatory of Japan (NAOJ), Mitaka, Tokyo 181-8588, Japan}
 \affiliation{KAGRA Observatory, Institute for Cosmic Ray Research (ICRR), University of Tokyo, Hida, Gifu 506-1205, Japan}
\author{Ayaka Shoda}
 \affiliation{National Astronomical Observatory of Japan (NAOJ), Mitaka, Tokyo 181-8588, Japan}
\author{Yoshinori Fujii}
 \affiliation{Department of Astronomy, University of Tokyo, Bunkyo, Tokyo 113-0033, Japan}
\author{Koki Okutomi}
\affiliation{KAGRA Observatory, Institute for Cosmic Ray Research (ICRR), University of Tokyo, Hida, Gifu 506-1205, Japan}
\author{Mark Andrew Barton}
 \affiliation{National Astronomical Observatory of Japan (NAOJ), Mitaka, Tokyo 181-8588, Japan}
 \affiliation{Institute for Gravitational Research, University of Glasgow, Glasgow G12 8QQ, UK}
\author{Ryutaro Takahashi}
 \affiliation{National Astronomical Observatory of Japan (NAOJ), Mitaka, Tokyo 181-8588, Japan}
\author{Kentaro Komori}
 \affiliation{Department of Physics, University of Tokyo, Bunkyo, Tokyo 113-0033, Japan}
\author{Naoki Aritomi}
 \affiliation{Department of Physics, University of Tokyo, Bunkyo, Tokyo 113-0033, Japan}
\author{Tomofumi Shimoda}
 \affiliation{Department of Physics, University of Tokyo, Bunkyo, Tokyo 113-0033, Japan}
\author{Satoru Takano}
 \affiliation{Department of Physics, University of Tokyo, Bunkyo, Tokyo 113-0033, Japan}
\author{Hiroki Takeda}
 \affiliation{Department of Physics, University of Tokyo, Bunkyo, Tokyo 113-0033, Japan}
\author{Enzo Nicolas Tapia San Martin}
 \affiliation{National Astronomical Observatory of Japan (NAOJ), Mitaka, Tokyo 181-8588, Japan}
 \affiliation{Nikhef, Science Park 105, 1098 XG Amsterdam, The Netherlands} 
\author{Ryohei Kozu}
 \affiliation{KAGRA Observatory, Institute for Cosmic Ray Research (ICRR), University of Tokyo, Hida, Gifu 506-1205, Japan}
\author{Bungo Ikenoue}
 \affiliation{National Astronomical Observatory of Japan (NAOJ), Mitaka, Tokyo 181-8588, Japan}
\author{Yoshiyuki Obuchi}
 \affiliation{National Astronomical Observatory of Japan (NAOJ), Mitaka, Tokyo 181-8588, Japan}
\author{Mitsuhiro Fukushima}
 \affiliation{National Astronomical Observatory of Japan (NAOJ), Mitaka, Tokyo 181-8588, Japan}
\author{Yoichi Aso}
 \affiliation{National Astronomical Observatory of Japan (NAOJ), Mitaka, Tokyo 181-8588, Japan}
  \affiliation{The Graduate University for Advanced Studies (Sokendai), Mitaka, Tokyo 181-8588, Japan}
\author{Yuta Michimura}
  \affiliation{Department of Physics, University of Tokyo, Bunkyo, Tokyo 113-0033, Japan}
\author{Osamu Miyakawa}
 \affiliation{KAGRA Observatory, Institute for Cosmic Ray Research (ICRR), University of Tokyo, Hida, Gifu 506-1205, Japan}
\author{Masahiro Kamiizumi}
 \affiliation{KAGRA Observatory, Institute for Cosmic Ray Research (ICRR), University of Tokyo, Hida, Gifu 506-1205, Japan}

\date{\today}

\begin{abstract}
This paper reports on the design and characteristics of a compact module integrating an optical displacement sensor and an electromagnetic actuator for use with vibration-isolation systems installed in KAGRA, the 3-km baseline gravitational-wave detector in Japan. 
In technical concept, the module belongs to a family tree of similar modules called OSEMs, used in other interferometric gravitational-wave detector projects. 
After the initial test run of KAGRA in 2016, the sensor part, which is a type of slot sensor, was modified by increasing the spacing of the slot from 5~mm to 15~mm to avoid the risk of mechanical interference with the sensor flag.
We confirm the sensor performance is comparable to that of the previous design despite the modification. We also confirm the sensor noise is consistent with the theoretical noise budget. The noise level is 0.5~nm/Hz$^{1/2}$ at 1~Hz and 0.1~nm/Hz$^{1/2}$ at 10~Hz, and the linear range of the sensor is 0.7~mm or more. We measured the response of the actuator to be 1~N/A, and also measured the resistances and inductances of coils of the actuators to confirm consistency with theory. Coupling coefficients among the different degrees of freedom were also measured and shown to be negligible, varying little between designs.
A potential concern about thermal noise contribution due to eddy current loss is discussed. 
As of 2020, 42 of the modules are in operation at the site. 
\end{abstract}

\maketitle

\section{Introduction}
Vibration isolation is an essential technique for state of the art instruments including gravitational-wave (GW) detectors to achieve accurate measurements. The terrestrial GW detectors such as LIGO~\cite{LIGO:2015}, Virgo~\cite{Virgo:2014}, and KAGRA~\cite{KAGRA:2018,Michimura:2019} are large-scale laser interferometers with baseline lengths of 3-4~km. Mirrors used in the interferometers are suspended by multi-stage pendulums in ultra-high vacuum for vibration isolation. 
The suspensions filter out seismic motion, reducing the motion transferred to the mirrors above the mechanical resonant frequencies of the pendulums.
For accurate observation of GWs with spacetime strain of $\sim 10^{-21}$, the vibration of the ground must be attenuated $10^9$ times or more above 10~Hz~\cite{Michimura:2017}.

In practice, we must also damp the oscillations of the pendulums, which amplify the seismic motion at the resonant frequencies. Kinetic energy stored in the resonant modes can be dissipated with active or passive damping control, but the damping control should not degrade vibration-isolation performance of the pendulums in the frequency range above 10Hz that is used for GW observation.
In this paper, we report on a compact module for such sophisticated active damping of the pendulums used in the KAGRA interferometer.

Technically, our module belongs to a family tree of designs 
combining optical sensors and electromagnetic actuators (OSEMs). In the late 1990s, one of the first generation of OSEMs was designed for LIGO~\cite{Kawamura:1997}. Since then, several types of OSEMs have been developed and incorporated into laser interferometric GW detectors~\cite{Carbone:2012,Moore:2014,Aston:2014,Fabian:2016,Shoda:2019}. 
Like previous designs, our OSEM integrates a contactless sensor and a contactless actuator into a fist-sized body that can be attached to a suspended damper.

An advantage of such a compact sensor-actuator is that it can push or pull on exactly the same point where the motion is sensed. This feature is helpful for designing control filters for damping. Regardless of the number of OSEMs attached to a suspended body, we can start with simple control filters to stabilize this suspension before we characterize sensors and actuators in situ to correct differences due to manufacturing and installation tolerances. Then we can implement more sophisticated control technology~\footnote{Private communication from Yutaro Enomoto and Masayuki Nakano, interferometer commissioners of KAGRA}.

\begin{figure}
\begin{center}
\includegraphics[width=8cm]{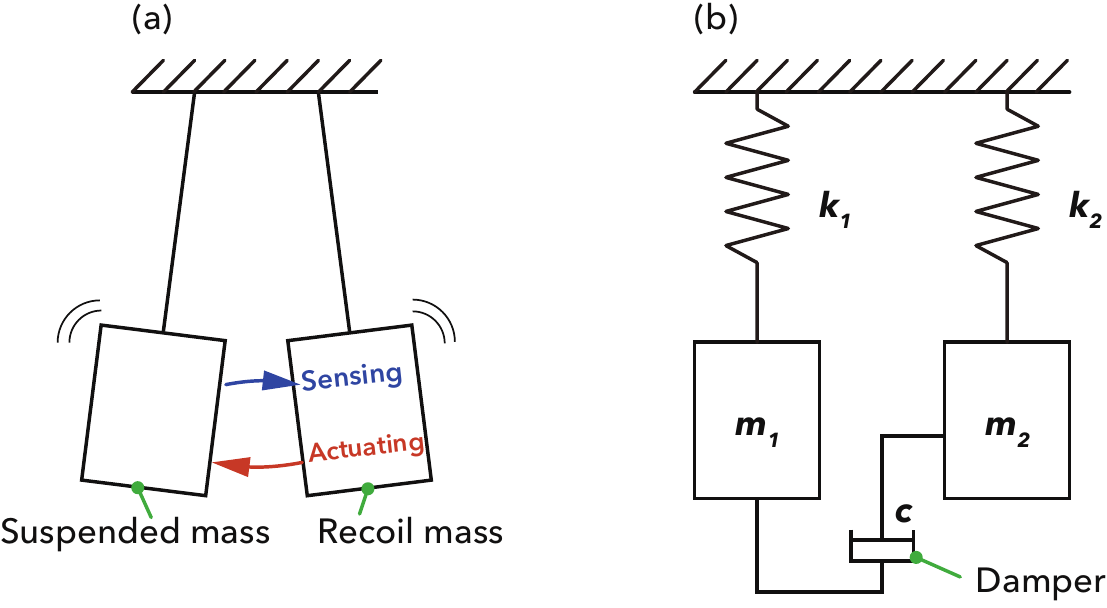}
\caption{Schematic view of a suspended mass paired with a suspended recoil mass having a sensor and actuator; (a) a cartoon of the system and (b) the dynamic model. Combined with a suitable control filter, the sensor and the actuator will work as a local damper for the suspended mass. In (b), $m_1$ and $m_2$ are masses of the suspended mass and the recoil mass, respectively; $k_1$ and $k_2$ are the spring constants of the respective suspensions; $c$ is the damping coefficient for the damper.}
\label{concept0}
\end{center}
\end{figure}

As an example of the suspended damper, let us consider a single pendulum with a recoil mass (Fig.~\ref{concept0}).
\begin{figure}
\begin{center}
\includegraphics[width=8cm]{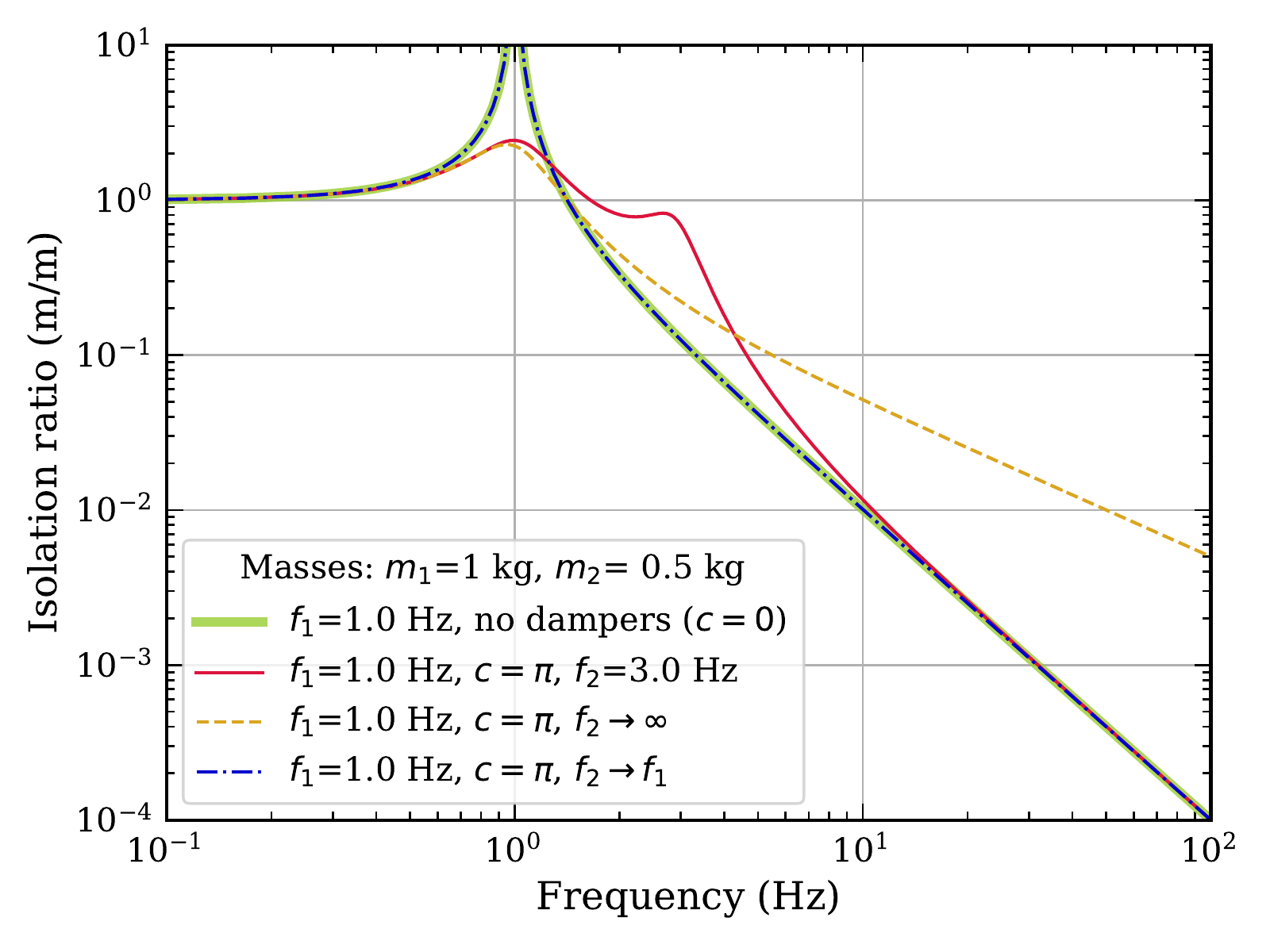}
\caption{Vibration isolation ratios with various local damping setups in the frequency domain for the suspended mirror shown in Fig.~\ref{concept0}. The thick green line is the ratio without damping. The red line shows a typical example with the suspended damper; the ratio converges to the thick green line above 10 Hz, while the resonant peaks are well suppressed. The other lines show two extreme cases. The dashed yellow line is for the case that the damper is fixed to the ground. The dash-dotted blue line is for the case where the masses have identical resonant frequency, and so are eventually undamped.}
\label{concept1}
\end{center}
\end{figure}%
In this case, the ``suspended mass'' represents an item such as a mirror to be isolated from seismic motion, and the OSEM is attached to the recoil mass.
Then, one can detect the displacement of the mirror with respect to the recoil mass using the sensor, and also damp the mirror motion using the actuator so that the mirror comes to be stationary with respect to the recoil mass. If the resonant frequency of the recoil mass pendulum, $f_2$, is appropriately shifted from that of the mirror pendulum, $f_1$, this system isolates the mirror from seismic motion without large low-frequency oscillations due to mechanical resonance of the suspension. This can be shown by simple calculations: the equations of motion of the system are
\begin{subequations}
\begin{eqnarray}
m_1 \ddot{x}_1 + c(\dot{x}_1-\dot{x}_2)+k_1(x_1-x_0)=0,\\
m_2 \ddot{x}_2 + c(\dot{x}_2-\dot{x}_1)+k_2(x_2-x_0)=0,
\end{eqnarray}\label{eqs:two-body_mech}
\end{subequations}%
where $x_0$ represents the seismic motion; $x_1$ and $x_2$ are the displacement of the mirror and the recoil mass, respectively; $m_1$ and $m_2$ are masses of the mirror and the recoil mass; $k_1$ and $k_2$ are spring constants of the respective suspensions; $c$ is the damping coefficient for the damper.
Here, we want to know the vibration-isolation ratio for the mirror, $x_1/x_0$. In the frequency domain, the equations of motion become
\begin{subequations}
\begin{eqnarray}
-\omega^2m_1 \tilde{x}_1 + ic\omega(\tilde{x}_1-\tilde{x}_2)+k_1\tilde{x}_1=k_1 \tilde{x}_0,\\
-\omega^2m_2 \tilde{x}_2 + ic\omega(\tilde{x}_2-\tilde{x}_1)+k_2\tilde{x}_2=k_2 \tilde{x}_0,
\end{eqnarray}
\end{subequations}
where $\omega$ is angular frequency, $\tilde{x}_p$ represents frequency components corresponding to $x_p$ for $p=0,1,2$, and $i$ is the imaginary unit. Solving the equations, one can obtain the ratio
\begin{equation}
\frac{\tilde{x}_1}{\tilde{x}_0}=\frac{1+2i(\xi_1\frac{\omega}{\omega_1}+\xi_2\frac{\omega}{\omega_2})-\frac{\omega^2}{\omega_2^2}}{(1+2i\xi_1\frac{\omega}{\omega_1}-\frac{\omega^2}{\omega_1^2})(1+2i\xi_2\frac{\omega}{\omega_2}-\frac{\omega^2}{\omega_2^2})+4\xi_1\xi_2\frac{\omega^2}{\omega_2\omega_1}}\label{eq:ratio},
\end{equation}
where $\omega_p\equiv\sqrt{k_p/m_p}=2\pi f_p$ and $\xi_p\equiv c/(2\sqrt{m_pk_p})$ for $p=$ 1 and 2, respectively.

Fig.~\ref{concept1} shows some examples of the form of Eq.~(\ref{eq:ratio}) as a function of temporal frequency, $f=\omega/(2\pi)$, for different parameters; here $m_1$ = 1~kg, $m_2$ = 0.5~kg, and $f_1$ = 1~Hz are fixed, while $c$ and $f_2$ are varied.
The thick green line is a reference case of $c=0$ or $\xi_1=\xi_2=0$, which corresponds to a simple suspended mirror without damping; the isolation ratio is proportional to $f^{-2}$ above $f_1$, while the mirror oscillates considerably at $f_1$. In contrast, the red line comes from a typical application of the suspended damper; it engages damping control of $c=\pi$ with the pendulum for the recoil-mass suspension, which is more rigid ($f_2$ = 3~Hz) than that for the mirror; $(\xi_1, \xi_2) = (1/4, 1/6)$. Compared to the reference line, the damping suppresses the peak around $f_1$ without degrading the frequency response above 10~Hz, although it introduces a small resonant enhancement of motion around $f_2$. 

The frequency response above 10~Hz will degrade with increasing rigidness of the recoil-mass suspension. The dashed yellow line in Fig.~\ref{concept1} shows an extreme case of where the damper is fixed to the ground. This is why we prefer the compact local sensor-actuator attachable to the suspended recoil mass. In reality, sensors fixed to the ground also take part in the damping control along with frequency-dependent filters.

Another extreme case is of $f_1=f_2$, where the denominator of Eq.~(\ref{eq:ratio}) becomes 0 when $f=f_1=f_2$. As shown by the dash-dotted blue line in Fig.~\ref{concept1}, the resonant peak diverges, and so the oscillation seen from the ground will continue forever in this model (which contains no natural damping), this is not useful as it means there is no damping of the resonant motion of the pendulum.

In this paper, we will report on the latest version of our OSEMs; we have designed and fabricated three versions of the modules within KAGRA (see Table~\ref{tab:versions}).
As of 2020, 42 OSEMs (ver.~2 in Table~\ref{tab:versions}) are used in the KAGRA interferometer. There are seven suspensions for room-temperature mirrors, in each of which six OSEMs are attached to the recoil mass for the intermediate mass. The intermediate mass is located at the second bottom level of the multi-stage pendulum system, and suspends the mirror at the lowest level~\cite{Shoda:2019}. 
The original concept was for there to be an additional four OSEMs on the recoil mass for the mirror, and this was adopted until the initial test run in 2016. 
To avoid the risk of mechanical interference (described in Section II for the detail), however, we did two things recently: (a) eliminating the sensor parts from four OSEMs on the mirror recoil mass, and (b) upgrading the OSEM design from ver. 1 to 2 by widening the sensor slot spacing.
Note that our OSEMs ver.~0 described in~\cite{Fabian:2016} were used in an off-site test and modified to ver.~1 for the initial test run by reducing the number of parts to make mass production within KAGRA easier.
\begin{table}
\caption{\label{tab:versions}Versions of the KAGRA OSEMs; in this paper, ver.~2 is mainly described. Comparisons with ver. 1 are also described.}
\begin{ruledtabular}
\begin{tabular}{cll}
Ver. &  Slot spacing & Explanations\\
\hline
0 & Narrow (5 mm)& For prototyping the KAGRA suspension~\cite{Fabian:2016}\\
1 & Narrow (5 mm)& Used until the initial test run in 2016\\
2 & Wide (15 mm)& In operation as of 2020 
\end{tabular}
\end{ruledtabular}
\end{table}%

In the remainder of this paper, we show the overall design of our OSEMs (ver.~2) in Section II, and describe the characteristics of the sensor and actuator parts in Section III and IV, respectively. Section V gives further detail of the design and issues with it. Section VI is a summary. Some useful theoretical formulae are summarized in appendices. Appendix A describes how to calculate the inductance of a multilayer solenoid coil.
Appendix B describes how to calculate the electromagnetic forces arising between a multilayer solenoid coil and a cylindrical magnet. Appendix C describes how to calculate the viscous damping coefficient due to eddy current loss in the coil bobbin of a multilayer solenoid.

\begin{figure*}
\begin{center}
\includegraphics[width=12cm]{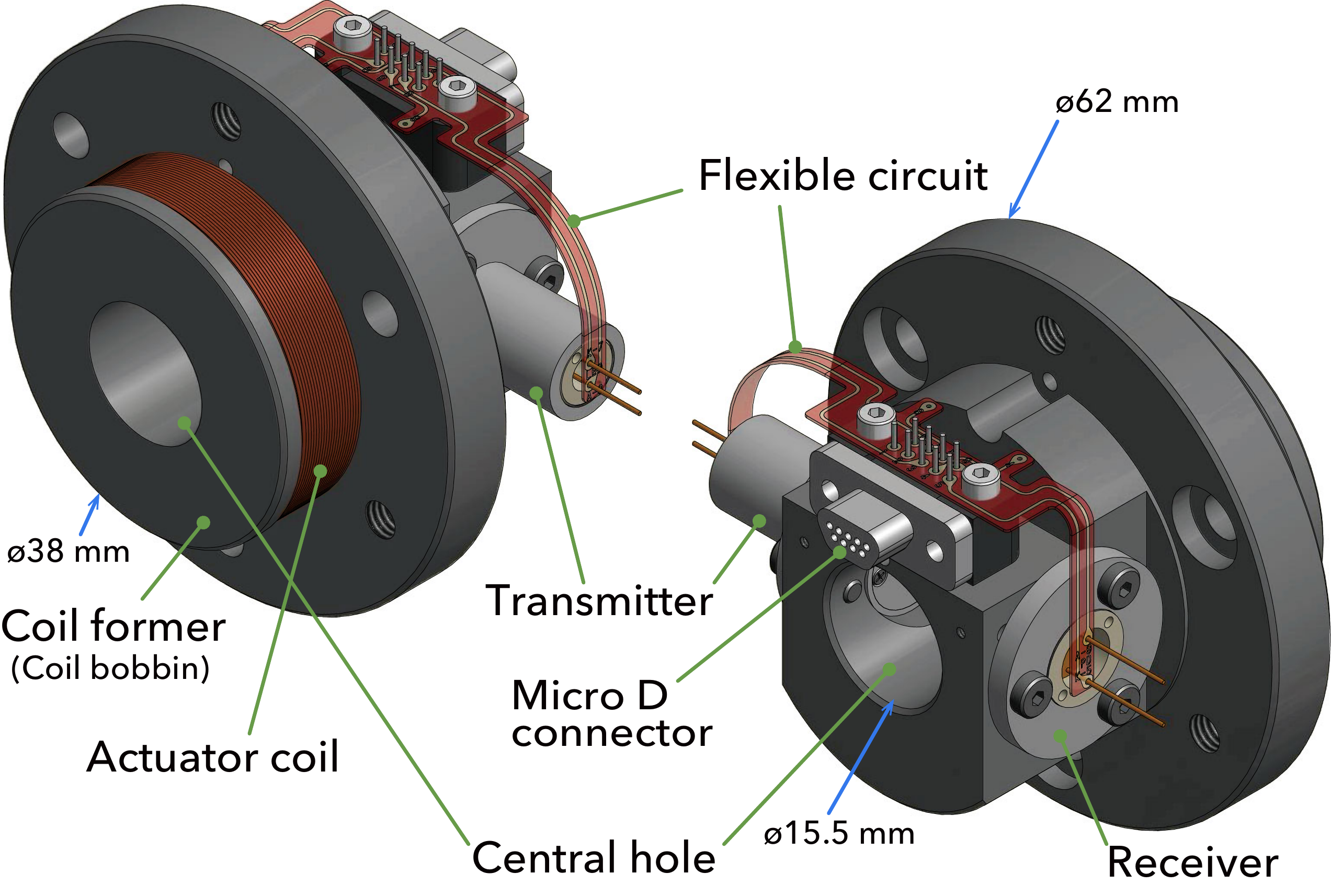}
\caption{Schematic view of a KAGRA OSEM (optical sensor and electromagnetic actuator) from two different viewpoints.}
\label{OSEM_all}
\end{center}
\end{figure*}%
\begin{figure*}
\begin{center}
\includegraphics[width=12cm]{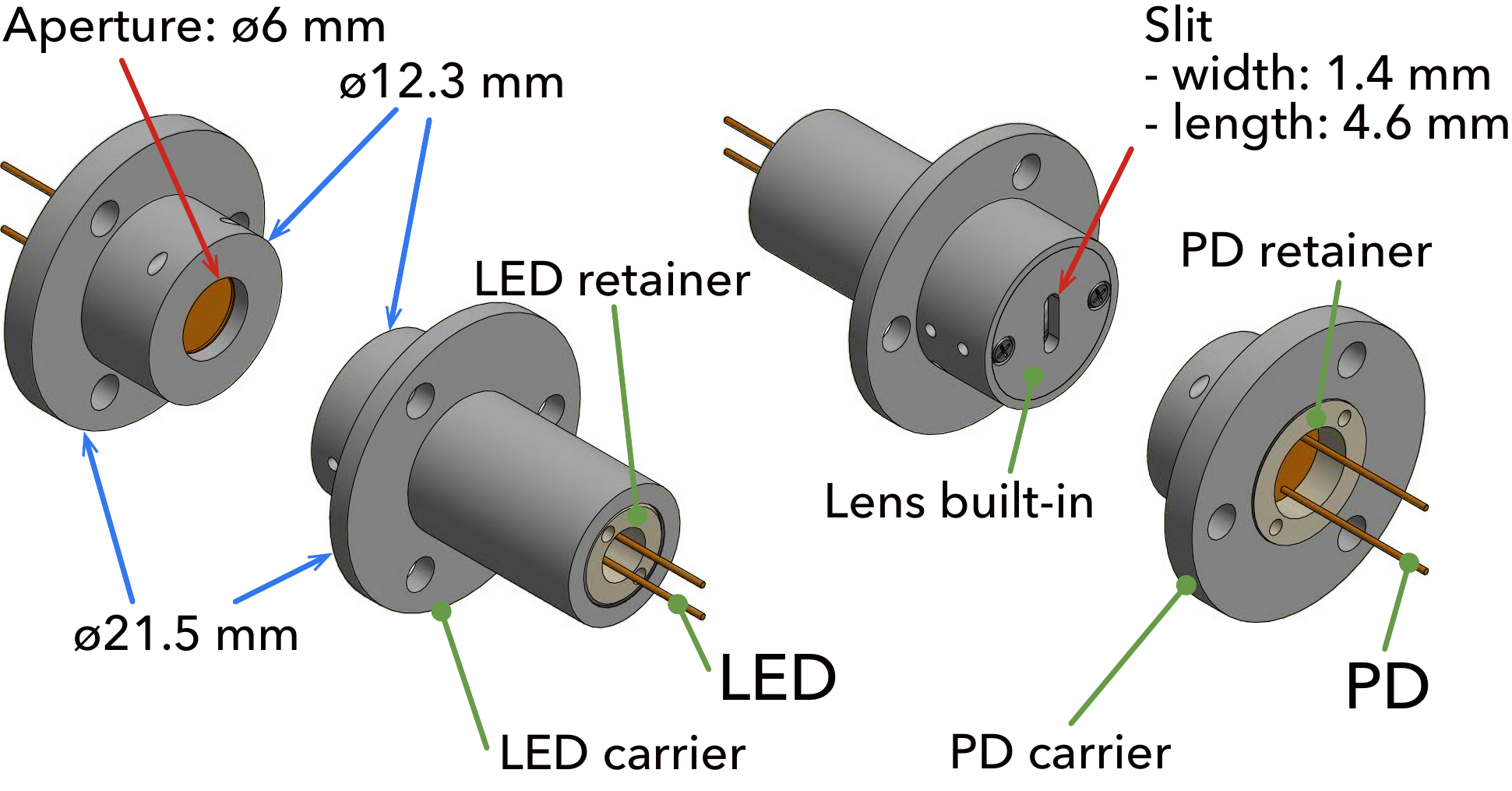}
\caption{Schematic view of the optical sensor in an OSEM from two different viewpoints; only the transmitter (TX) and the receiver (RX) appear from Fig.~\ref{OSEM_all}. The TX contains an LED, while the RX contains a PD. See also Fig.~\ref{OSEM_w_flag} for the cross-sectional view.}
\label{OSEM_sensor}
\end{center}
\end{figure*}%

\section{Design}
This section gives an overview of the design of the OSEMs currently installed at KAGRA, and more detailed descriptions of the sensor and actuator parts.
\subsection{Overview of the design}
Fig.~\ref{OSEM_all} shows a schematic view of our OSEM. An optical sensor and an electromagnetic coil are integrated into a single module.
The main body of the module is made of polyether ether ketone (PEEK), and the outer diameter is 62~mm. A bobbin structure protrudes from the main body to be a coil former, and has an outer diameter of 38~mm with a central through hole. The central through-hole is 15.5~mm in diameter and the main body is 42~mm long. 

\begin{figure*}
\begin{center}
\includegraphics[width=17cm]{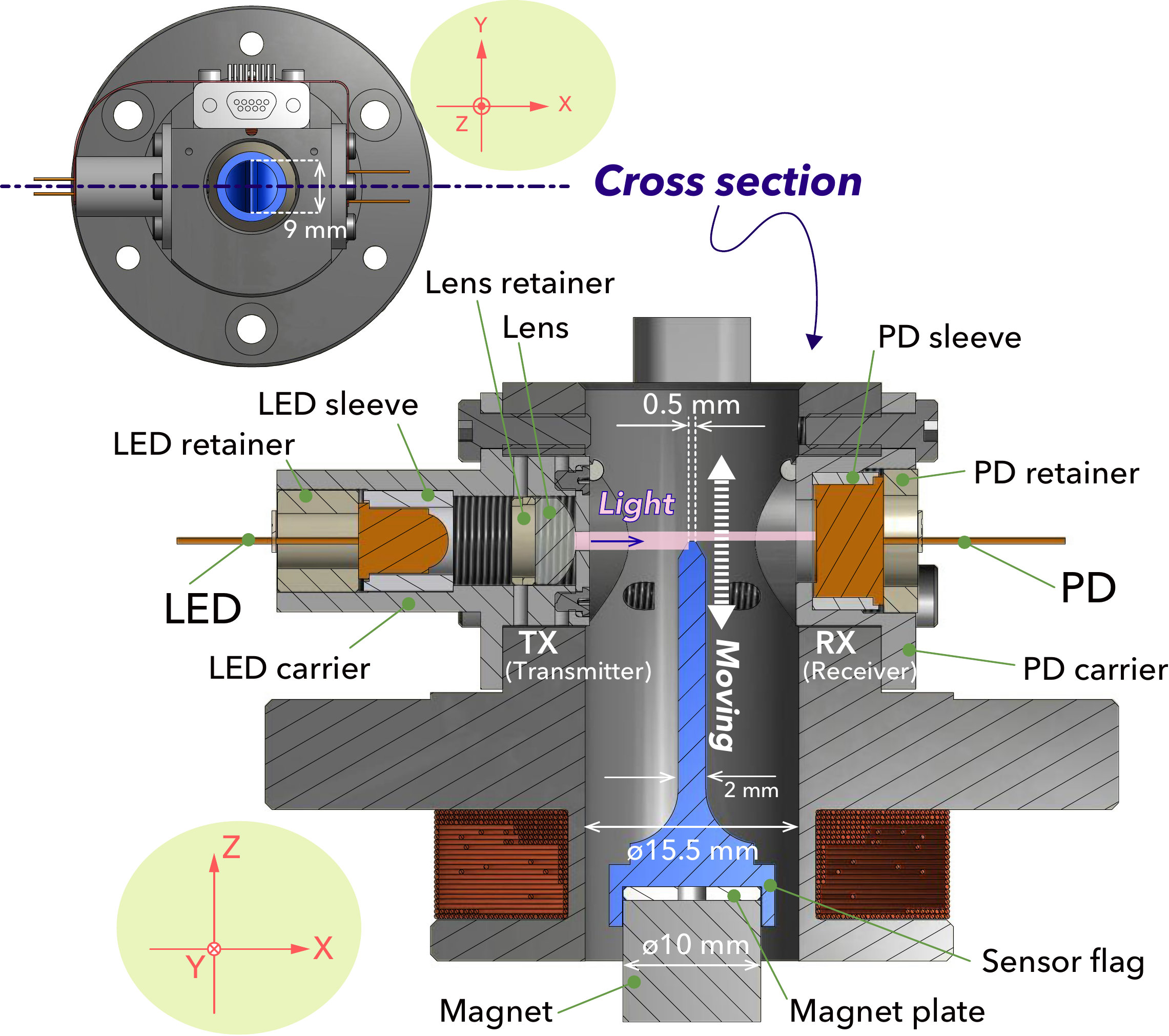}
\caption{Schematic rear view of the OSEM (upper left), and the cross-sectional view (bottom center). In order to relate directions, a local coordinate indicator is attached to each view.}
\label{OSEM_w_flag}
\end{center}
\end{figure*}

The optical sensor for displacement measurement is a shadow sensor consisting of a transmitter (TX) and a receiver (RX); see also Figs.~\ref{OSEM_sensor}~and~\ref{OSEM_w_flag}. A light-emitting diode (LED) in the TX emits a light beam, which crosses the central hole, and illuminates a photodioide (PD) in the RX. 
The tip of the sensor flag (see Fig.~\ref{OSEM_w_flag}) protrudes into the central hole, and partly shades the light beam.
The light power reaching the PD varies monotonically with respect to the insertion depth, as does the photocurrent from the PD. As a whole, this setup works as a displacement sensor for the relative position between the flag tip and the OSEM body.

The electromagnetic actuator is a coreless voice coil actuator without yoke, consisting of a cylindrical permanent magnet and the coil (Fig.~\ref{OSEM_w_flag}). 
The electric current carried in the coil wire generates an electromagnetic force between the OSEM body and the magnet.
The magnet is integrated into the sensor flag. The assembly of the sensor flag and the magnet is attached to the intermediate mass, while its counterpart, the OSEM body, is attached to the recoil mass for the intermediate mass~\cite{Fabian:2016,Shoda:2019}.

The electronic interface of the sensor and the actuator is a micro-D connector (plug/male; Glenair, GMR7590-9P-1BPN). 
The connector, the LED, the PD, and both ends of the coil wire are all soldered to a flexible printed circuit made of polyimide film for low outgassing rate
and for a reasonable production cost.
The thickness of the film is $50\,\mathrm{\mu m}$, and electronic lines are printed with copper foil of $35\,\mathrm{\mu m}$ in thickness and 0.4~mm in width.

In operation, the maximum rating of current for the coil is set at 100~mA~\cite{Michimura:2017} to prevent damage of the coil from Joule heating (taking into account the experimental investigations in LIGO~\cite{Barton:2010B,Aston:2014,Coyne:2014}). Note that the Joule heat cannot escape via convection in vacuum, where the OSEMs are used. Of the parts for the OSEM, the LED and PD are the weakest against heat; their maximum operating temperature is 100$^\circ$C. Aluminum alloy used for the holders of the LED or the PD has a recrystallization temperature of $\sim 150^\circ$C, and so we usually avoid heating the material above $\sim 120^\circ$C to maintain a safety margin, even during baking. A tiny amount of epoxy adhesive, Loctite Ablestik 2116 by Henkel, is used to fix the coil wire to the body, and has a maximum operating temperature of 130$^\circ$C.
As seen later, the coil resistance is $\sim 11\,\Omega$, and the current of 100~mA on the resistance will cause Joule heat of $\sim 0.1$~W.
No critical damage has been observed for the OSEMs in operation so far.
When one would like to modify the design further or increase the maximum rating of current for the coil, it would be nice to investigate the thermal designing combined with the rate of outgassing again.

\subsection{Sensor part}
The transmitter (TX) and the receiver (RX) work as a shadow sensor together with the sensor flag (Figs.~\ref{OSEM_sensor} and~\ref{OSEM_w_flag}). The fundamental design of the TX and RX follow those of the Birmingham OSEMs (BOSEMs) of LIGO~\cite{Aston:2011,Carbone:2012,Aston:2014}.

The major difference between the current design (ver.~2) and the previous one (ver.~1 or 0) is that the separation between the TX and RX has been widened (see Table~\ref{tab:versions}). In our OSEM ver.~0 or 1, the opposing surfaces of the ends of the TX and RX stuck out into the central hole of the OSEM body, and the separation of the surfaces was as narrow as 5~mm. As the sensor flag had a 2-mm thickness (Fig.~\ref{OSEM_w_flag}), there were only 1.5-mm gaps on each side of the flag when it was located at the nominal position. With this setup, there was a risk that the flag would touch and be knocked off by the TX or RX.
This was a particular problem for the mirror and the intermediate mass, where the flags were attached with EP30 adhesive. Such accidents occurred several times during installation and commissioning activities before the initial test run of KAGRA and also in prototype testing. To mitigate the risk of such mechanical interference, the TX and RX were retracted in a redesign such that their faces are almost along the inner surface of the central hole in the current design; see the cross sectional view in Fig.~\ref{OSEM_w_flag}. Currently, the separation is 15~mm (note that the inner diameter of the central hole itself is 15.5~mm), and the nominal gap around the flag is widened from 1.5~mm to 6.5~mm.
As a drawback, the wider gap reduces light power arriving at the PD by about half. The effects are discussed later in this paper.
As a further precaution, the flags for the intermediate recoil mass were changed to a magnetically self-assembling design; see the next section.

The TX encapsulates an LED in an aluminum structure, the LED carrier (Fig.~\ref{OSEM_sensor}), in which the LED is held by a sleeve made of machinable ceramic, Macor by Corning, for electric insulation. The LED is a TSTS7100 by Vishay in a TO-18 package, which emits light at 950~nm. An uncoated planoconvex lens made of BK7, 08PQ06 by Comar Optics, is put in front of the LED for collimating the emitting light beam (Fig.~\ref{OSEM_w_flag}).
The LED and the lens are fixed to the carrier with PEEK retainers.
A mask with a slit of width 1.4~mm and length 4.6~mm is attached in front of the lens to limit the light beam to the PD. The width is determined so that it gives a suitable linear range for the shadow sensor (Fig.~\ref{SensCurves}), while the length is determined to be slightly larger than the PD photosensitive surface, which is a 3.4-mm square.

The RX encapsulates a PD in an aluminum structure, the PD carrier (Fig.~\ref{OSEM_sensor}), in which the PD is held by a sleeve made of Macor (Fig.~\ref{OSEM_w_flag}). The PD is an S1223-01 by Hamamatsu Photonics in a TO-5 package, which has a response of about 0.58~A/W at the working wavelength of the LED, 950~nm~\footnote{We do not put an optical filter in front of the PD.}. The PD is fixed to the carrier with a PEEK retainer. The PD carrier has an aperture 6~mm in diameter, which corresponds to the input window of the PD (5.9~mm in diameter).
As seen in Section III, the linear range of the shadow sensor is about 1~mm, which is smaller than the photosensitive area. When one wants to downsize the sensor part, one may consider using a smaller photodiode such as a BPX65 by Centronic.

\subsection{Actuator part}
An electromagnetic coil serves as an actuator to exert a force between the the OSEM body and a magnet on the sensor flag (Fig.~\ref{OSEM_w_flag}). The permanent magnet, KE110 by Niroku, has a cylindrical shape 10~mm in diameter and 10~mm in thickness, and is made of samarium-cobalt (SmCo) YKG28 (which is the name of a material by Niroku) with nickel plating~\cite{Michimura:2017}. The magnet is attached on the bottom of the sensor flag, which is made of aluminum alloy for non-magnetization.

The electromagnetic coil is a multilayer coreless solenoid formed with a single copper wire with polyimide coating, of magnet wire type NW16-C of the National Electrical Manufacturers Association (NEMA). A bare conductor in the wire has a diameter of 0.32~mm, corresponding to AWG-28, while the total outer diameter including the coating is 0.36~mm (B1282803 by MWS Wire Industries). The wire is wound on a bobbin structure with $\sim 600$ turns in total, 22 turns per layer and 28 layers in design. The inner diameter and axial length of the coil are 18~mm and 8~mm, respectively. 

In practice, fabrication errors are unavoidable for coil winding. For example, 8~mm/0.36~mm $\simeq 22.2$ turns, so the number of turns per layer in design is an approximation. Similarly, the outer diameter of the coil differs from a naively calculated value of 38.16~mm (18~mm plus twice 28 layers of 0.36~mm circles). The circular cross sections can be more closely packed within a limited area in the manner that the distance of each layer will be $\sqrt{3}/2$ of each circle diameter. Then, the minimum outer diameter of the coil is 35.46~mm in theory. In fact, the measured value is about 36.5~mm, which is close to midway between the two extreme cases. In addition, the outermost layer of the winding sometimes finishes in the middle.

From the design, the inductance and resistance of the coil can be computed. Here, we assume 27.5 layers for the coil winding to take fabrication errors into account. For the inductance, 8.9~mH is obtained by the method described in Appendix~\ref{Sec:calc_induct}. The total length of the wire for coil winding is expected to be $\sim 53.0$~m from summing up circumferences of every wire loop. From this number, $11.3\,\Omega$ is obtained for the resistance, where we use a value of resistance per meter $0.214\,\mathrm{\Omega/m}$ at 20$^\circ$C found in a specification document of the wire.

The coil former or the bobbin structure is made of carbon-loaded PEEK, Ketron CA30 (the old name is PK-450CA) by Mitsubishi Chemical Advanced Materials, with conductivity chosen as a trade-off to prevent unwanted charge accumulation by triboelectric or other charging mechanisms in the body while avoiding unwanted loss of kinetic energy from eddy currents in the bulk, and unwanted magnetization of the body.
The specification document of this material, however, does not clearly show the lower limit of the electrical resistivity.
We look into this point further in the discussion section. We did not have a chance to change the coil formers, which are unchanged since our OSEM ver.~0~\cite{Fabian:2016}. In contrast, the coil wires were changed when upgraded to OSEM ver.~1 from thinner ones (0.20~mm diameter) used in the OSEM ver.~0 to optimize the control scheme~\cite{Michimura:2016,Michimura:2017}.

While redesigning the OSEM to ver.~2, we also modified the flags to be magnetically self-assembling, so as to reduce the risk of damage in the event of them being bumped.
The detail is out of the scope of this paper, and partly discussed in our internal document~\cite{Barton:2016}, so here we touch on it briefly. 
Instead of gluing the sensor flag directly to magnet, a thin SS400 steel disk (``magnet plate'' in Fig.~\ref{OSEM_w_flag}) is glued on the bottom of the flag.
The steel disk then magnetically adheres to the magnet; the same treatment is done for the base between the magnet and the intermediate mass. Before the modification, during the installation and operation of the suspensions, the magnets could sometimes knock against the surrounding structures and be detached at the bonding areas. Once that happened, the recovery would take a long time, typically about one week, for cleaning the remnant of the bond and re-gluing. 
Thanks to the modification, we can easily reattach the flag tips and magnets even when they are accidentally knocked.
When such an incident occurs while assembling the suspension, we do a quick check of this multi-stage suspension by measuring several transfer functions among each different stage with local sensors (including the OSEMs under suspicion). Due to this quick check, we can learn whether the incident would affect the suspension performance by comparing it with the measured transfer functions in the past; if required, we can easily adjust the position of the OSEM body to recover the set point for the reattached flag tip.

In principle, the flag may be knocked off inside the vacuum chamber during an earthquake and snapped back in a slightly different place. There were several large earthquakes in the past, and every time that happened, we repeated the quick check introduced above. As a result, we did not have to open the chamber to reset the OSEM position, certainly after we changed the design of the sensor part by widening the separation between the transmitter and receiver units. In this suspension, we also have structures to prevent much displacement of suspended stages, and thanks to this widened separation, these structures also prevent direct mechanical interference between the flag and the OSEM body.

\section{Characterizing the sensors}
This section reports on the gain and the noise of the sensor part of the OSEMs. Note that \textit{sensitivity} is a word widely used with different meanings; one is a ratio of the output with respect to the input, and another is like a minimum threshold value of the input above which the input apparently changes the output. Therefore, we will use simply gain and noise, respectively, to avoid the confusion.

\subsection{Sensor gain}
The gain of the sensor can be derived from the response of the sensor as a function of flag position.
\begin{figure*}
\begin{center}
\includegraphics[width=17.5cm]{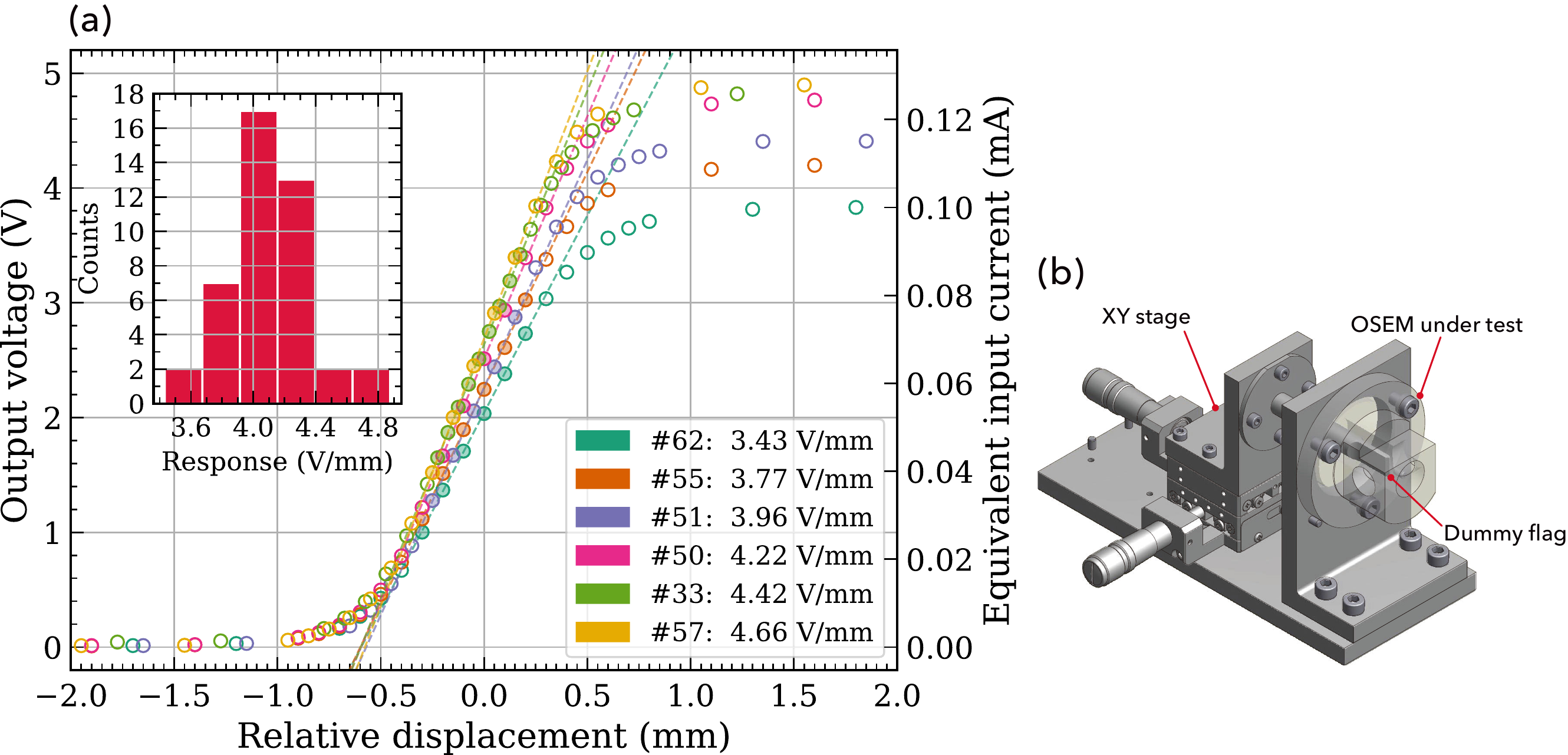}
\caption{(a) Typical measured responses of the sensors with respect to the relative displacement of the tip of the dummy sensor flag. The vertical axis on the left indicates single-ended output voltage from a transimpedance amplifier in the driver circuit, while it is converted (divided by 38.3~k$\Omega$) to the equivalent photocurrent into the circuit on the right. The measurements are shown by circles; the data subsets shown by filled circles are used for linear fit to estimate the sensor gains (V/mm in this figure); the dashed lines are the fit curves. The inset histogram shows the distribution of the gains of all the fabricated sensors. All the estimated gains are listed in Table~\ref{tab:responses}.
(b) Schematic view of a clean-room compatible testbench for the OSEMs. The dummy sensor flag is attached onto the XY translation stage and movable with two micrometers.}
\label{SensCurves}
\end{center}
\end{figure*}%
Before shipping OSEMs to the KAGRA site, we measured the responses one by one with a testbench in a clean booth (Fig.~\ref{SensCurves}).
\begin{table}
\caption{\label{tab:responses}Estimates of the gains of the sensors, and measurements of the resistances and inductances of the actuator coils. The serial numbers (S/Ns) are for identification of each OSEM.}
\begin{ruledtabular}
\begin{tabular}{cccc}
S/N & Gain (V/mm) & R ($\Omega$) & L (mH) \\
\hline
22 & $4.153\pm 0.034$ & 11.33 & 8.40\\
23 & $   3.800\pm    0.029$ &    11.58 &     8.40\\
24 & $   4.340\pm    0.021$ &    11.30 &     8.40\\
25 & $   4.876\pm    0.015$ &    11.68 &     8.40\\
26 & $   4.383\pm    0.060$ &    11.66 &     8.44\\
27 & $   3.983\pm    0.031$ &    11.33 &     8.40\\
28 & $   3.937\pm    0.053$ &    11.38 &     8.45\\
29 & $   4.367\pm    0.032$ &    11.22 &     8.47\\
30 & $   4.173\pm    0.032$ &    11.34 &     8.45\\
31 & $   4.007\pm    0.029$ &    11.25 &     8.36\\
32 & $   4.247\pm    0.056$ &      (no data) &     (no data)\\
33 & $   4.420\pm    0.027$ &    11.37 &     8.45\\
34 & $   4.097\pm    0.053$ &    11.45 &     8.37\\
35 & $   4.510\pm    0.032$ &    11.20 &     8.43\\
36 & $   4.100\pm    0.029$ &    11.17 &     8.39\\
37 & $   4.330\pm    0.051$ &    11.52 &     8.43\\
38 & $   4.098\pm    0.046$ &    11.23 &     8.47\\
39 & $   4.029\pm    0.033$ &    11.33 &     8.41\\
40 & $   4.005\pm    0.048$ &    11.31 &     8.41\\
41 & $   4.085\pm    0.029$ &    11.31 &     8.42\\
42 & $   4.363\pm    0.019$ &    11.29 &     8.45\\
43 & $   4.027\pm    0.026$ &    11.36 &     8.46\\
44 & $   3.957\pm    0.048$ &    11.31 &     8.48\\
45 & $   4.078\pm    0.040$ &    11.21 &     8.50\\
46 & $   4.390\pm    0.023$ &    11.30 &     8.10\\
47 & $   4.288\pm    0.034$ &    11.60 &     8.50\\
48 & $   3.811\pm    0.099$ &    11.35 &     8.40\\
49 & $   4.225\pm    0.051$ &    11.61 &     8.50\\
50 & $   4.218\pm    0.033$ &    11.28 &     8.50\\
51 & $   3.965\pm    0.051$ &    11.33 &     8.50\\
52 & $   3.438\pm    0.016$ &    11.63 &     8.40\\
53 & $   4.071\pm    0.031$ &    11.28 &     8.37\\
54 & $   4.008\pm    0.082$ &    11.62 &     8.39\\
55 & $   3.766\pm    0.042$ &    11.63 &     8.38\\
56 & $   4.301\pm    0.055$ &    11.31 &     8.38\\
57 & $   4.659\pm    0.036$ &    11.54 &     8.42\\
58 & $   3.723\pm    0.060$ &    11.58 &     8.42\\
59 & $   4.040\pm    0.087$ &    11.58 &     8.35\\
60 & $   3.730\pm    0.028$ &    11.51 &     8.32\\
61 & $   3.732\pm    0.035$ &    11.57 &     8.36\\
62 & $   3.433\pm    0.025$ &    11.57 &     8.37\\
63 & $   3.799\pm    0.046$ &    11.56 &     8.27\\
64 & $   4.321\pm    0.008$ &    11.60 &     8.35\\
\end{tabular}
\end{ruledtabular}
\end{table}%
The testbench has a dummy sensor flag (Fig.~\ref{SensCurves} (b)), with the same shape as the actual sensor flag shown in Fig.~\ref{OSEM_w_flag}. The dummy flag is attached onto an XY translation stage with micrometers. We can insert or retract the dummy flag into or out of the OSEM under test by adjusting the micrometers.

Typical responses are shown in Fig.~{\ref{SensCurves}} (a) for six samples out of the 43 units in total.
The measurements are shown by empty or filled circles. The dashed lines indicate linear fits to the measurements, where the data subsets used for the fitting are shown by the filled circles. The estimated slopes or the gains are also shown in the legend along with the serial numbers assigned to each OSEM for identification. The vertical axis in the left indicates single-ended output voltage from a driver circuit for the sensors. The driver circuit of the sensor for the measurements was the same type as used for KAGRA. Note that the driver circuits are usually used in differential signaling at the site, but our measurements were done single-ended for simplicity. For reference, equivalent photocurrent input to a transimpedance stage in the driver circuit is shown in the right vertical axis. The transimpedance is 38.3~k$\Omega$, so the photocurrent can be obtained by dividing the measured voltages by this resistance.

The overall shape of the response curve is similar for all the sensors, except for a vertical scale factor. The displacement range of the dummy flag in which the response is linear, or linear range, is constant at about 0.7~mm, while the gain is proportional to the maximum photocurrent that is obtained when the sensor flag does not occult the light beam at all. 

All the estimated gains are listed in Table~\ref{tab:responses}, and a distribution histogram of them is shown as an inset in Fig~\ref{SensCurves} (a). For the gains, the mean and the standard deviation are 4.09 and 0.288~V/mm, respectively, while the median is 4.08~V/mm. Note that the mean or median corresponds to $\sim 0.106$~A/m in terms of the photocurrent. The distribution has a peak around the mean or median. The six data sets plotted as typical responses in Fig~\ref{SensCurves} are randomly sampled from each bin of the histogram.
Currently, 42 of these OSEMs are installed in KAGRA.

\subsection{Sensor noise}
\begin{figure*}
\begin{center}
\includegraphics[width=17cm]{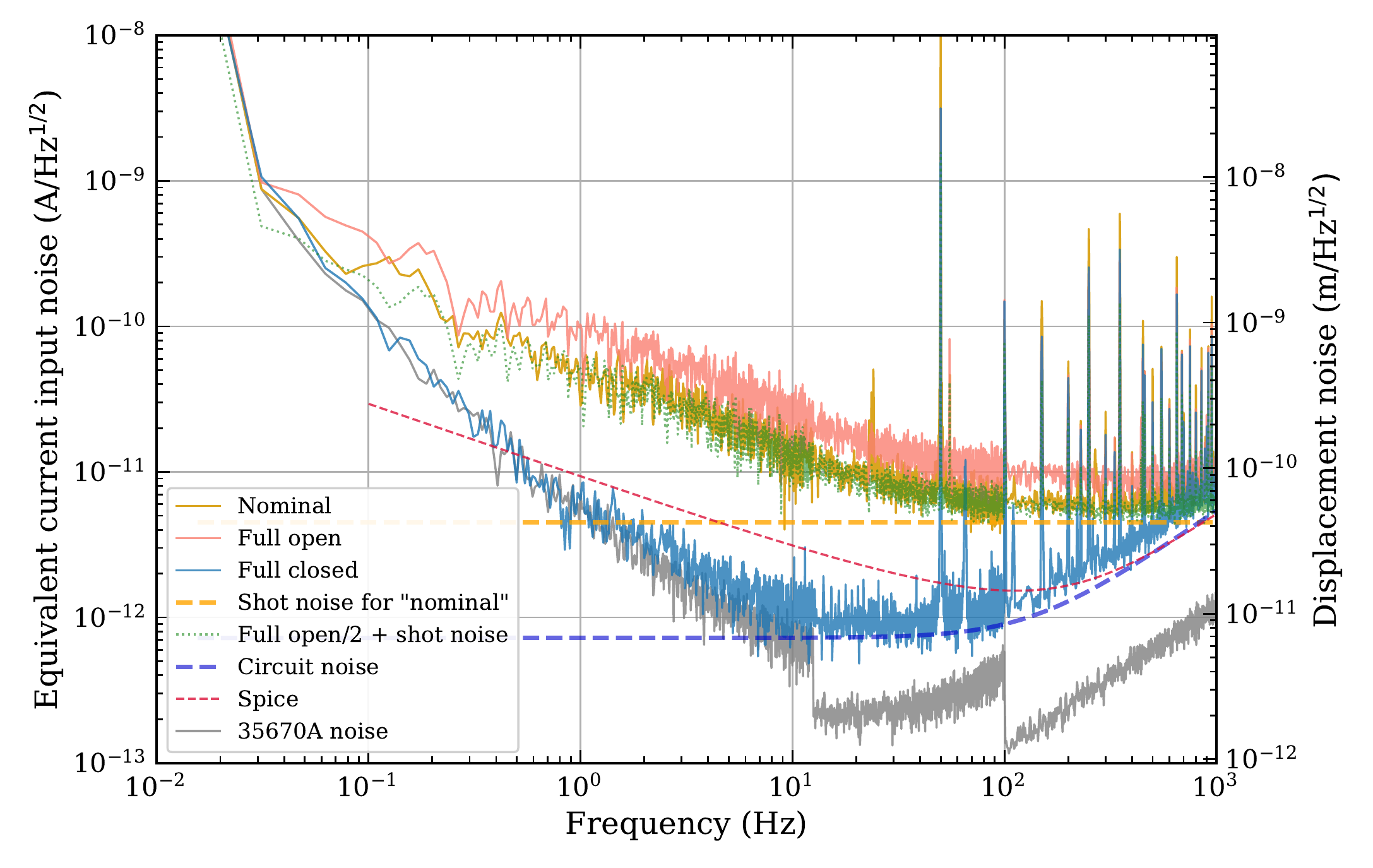}
\caption{One-sided amplitude spectral density of the sensor noise, and the relevant several measurements and theoretical lines. The vertical axis on the left is for showing the data in terms of the equivalent current input noise at the transimpedance stage of the sensor driver circuit, while they are converted to the output noise of the displacement sensor on the right.}
\label{fig:noisebudget}
\end{center}
\end{figure*}

To estimate the sensor noise, we measured and analyzed the noise of the sensor in a spare OSEM left off site, which was not installed at KAGRA. Fig.~\ref{fig:noisebudget}
shows one-sided amplitude spectra of several relevant measurements and some theoretical curves for the analysis. The vertical axis on the left of the figure indicates equivalent current input at the transimpedance opamp of the driver circuit, while it is converted to displacement noise on the right. The conversion factor (or gain) was measured as $\sim 0.106\,\mathrm{A/m}$ in the same manner as described in the previous subsection. All measurements in the figure were done in air, but the testbench was covered with an opaque shield to isolate it from ambient light and sound.

The solid orange curve ``nominal'' in Fig.~\ref{fig:noisebudget} is a measurement corresponding to the noise output of the sensor when the sensor flag is fixed at the middle of the linear range of the response function, or at the nominal position. The spectrum indicates $0.48\,\mathrm{nm/Hz^{1/2}}$ at 1~Hz, and $0.13\,\mathrm{nm/Hz^{1/2}}$ at 10~Hz. Even though the separation between the TX and the RX is widened, and so the light power received by the PD decreases by half, the noise level is comparable to that of the LIGO BOSEMs~\cite{Carbone:2012,Aston:2011}.

The nominal noise level can be explained by contributions from the intensity noise of the light source (LED) and the shot noise for the nominal photocurrent.
The solid red curve ``full open'' in Fig.~\ref{fig:noisebudget} was measured when the sensor flag was extracted out of the OSEM central hole, so the photocurrent in the sensor output took the maximum value, or twice the nominal one; the nominal photocurrent can be estimated as $62.6\,\mathrm{\mu A}$ from the measured output voltage 2.4~V and the DC transimpedance of $38.3\,\mathrm{k\Omega}$ in the driver circuit. The shot noise for the nominal photocurrent is $4.5\,\mathrm{pA/Hz^{1/2}}$, and is drawn with the dashed orange line in the figure. The dotted green curve is the root-mean-square sum of the shot noise and half of the ``full open'' curve (here its shot noise subtracted beforehand), and matches the nominal noise level very well.

The sensor output is dominated by the shot noise above 100~Hz. Below that frequency, it is dominated by the intensity noise of the LED down to $\sim 0.1$~Hz. In the interferometer, these sensors will only be used for detecting mechanical resonances in the low-frequency region.
Because both the intensity noise and the sensor gain will increase linearly with increasing light power on the PD, the resultant calibrated sensor noise will not be improved in the low-frequency region by simply increasing the light power.

The solid blue curve ``full closed'' in Fig.~\ref{fig:noisebudget} was measured when the sensor flag was inserted into the OSEM central hole well past the working point so that the PD was totally shaded by the sensor flag. The measurement corresponds to the noise contribution of the driver circuit. The dashed blue line shows a theoretical estimate of the circuit noise. The calculation includes noise contributions from the thermal noise of the transimpedance ($38.3\,\mathrm{k\Omega}$ and $100\,\mathrm{nF}$ in parallel), and the voltage and current noise of the opamp, which is an OP2177 by Analog Devices. Although the theoretical line is slightly lower than the measurement, it can explain the approximate magnitude and frequency dependence of the measurement above 10~Hz. Below that frequency, the measurement is contaminated by the noise from the measuring instrument, an Agilent 35670A (shown in the grey line). Above 100~Hz, the theoretical noise is dominated by the opamp noise, while thermal noise of the transimpedance dominates below the frequency.

For comparison and confirmation, we also ran a circuit simulator, LTspice by Linear Technology, to estimate the noise contribution of the driver circuit; see the dashed red line in Fig.~\ref{fig:noisebudget}. Unfortunately, the corner frequency of the flicker noise appears to be different between the spice model of the opamp provided by the company and its specification document. The discrepancy between the simulator result and the measurement (the solid blue curve) would happen due to this issue. According to the specification document, the corner frequency should be around 3~Hz, while the opamp Spice model shows it around 50~Hz. Note that, for simplicity, the flicker-noise behavior was not taken into account when calculating the theoretical noise shown in the dashed blue line.

If one would like to improve the noise level more at 1~Hz, where the contribution from the LED intensity noise is dominant, the intensity noise should be reduced. One way would be to survey low-noise LEDs and pick a suitable one. The other way would be to modify the LED driver circuit to include intensity-noise stabilization. Another way to reduce the intensity noise would be to use a two-PD differential sensing scheme~\cite{Dumas:2009}.
According to the noise budget in Fig.~\ref{fig:noisebudget}, there should be room to improve by about one order of magnitude at 1~Hz.

\section{Characterizing the actuators}
This section reports on the actuator part of the OSEMs, especially the resistances and inductances of the coils, and the electromagnetic forces.

\subsection{Resistance and inductance of the coils}
\begin{figure}
\begin{center}
\includegraphics[width=8.6cm]{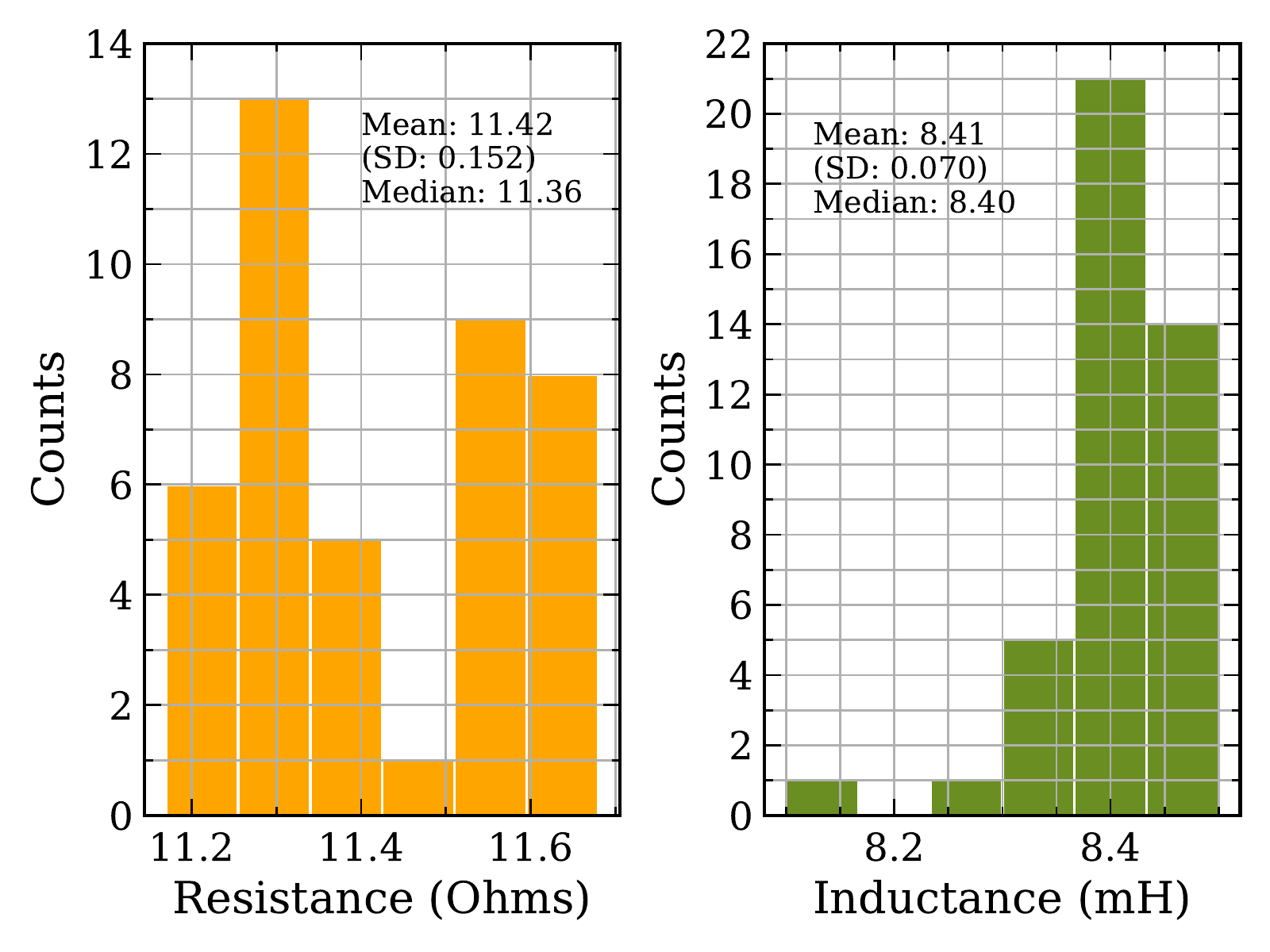}
\caption{Distributions of measured resistance and inductance of the OSEM actuator coils; the mean, standard deviation (SD), and median are also shown. With this binning, the inductance distribution shows a concentration at around 8.4 mH, while the resistance shows two peaks at around 11.3 $\Omega$ and 11.6 $\Omega$. The raw data for the histograms are listed in Table~\ref{tab:responses}.}
\label{R_and_L}
\end{center}
\end{figure}

Before shipping OSEMs to the KAGRA site, we measured resistances and inductances of the electromagnetic coils for the actuators; see Table~\ref{tab:responses}. Distributions of them are summarized in Fig.~\ref{R_and_L}.
In each histogram, the total number of the samples is 42.

For resistance, the mean of the measured values is $11.42\,\Omega$ with a standard deviation of $0.152\,\Omega$, but the measured values are not concentrated around the mean as shown in the histogram (the left in Fig.~\ref{R_and_L}), where two peaks are found at $11.3\,\Omega$ and $11.6\,\Omega$. The median is $11.36~\Omega$. For inductance, the mean of the measured values is 8.41~mH with a standard deviation of 0.070~mH. The median is 8.40~mH, which is equivalent to the mean. The single peak in the histogram (the right in Fig.~\ref{R_and_L}) corresponds to the mean and median. In summary, the measured values fall within a few percent of the mean values for both resistance and inductance.

As mentioned in Section II, the design gives a resistance of 11.3~$\Omega$ and an inductance of 8.9~mH. The estimated resistance and inductance agree with our measurements. For resistance, the designed value falls within the standard deviation of the measurements, and is close to the median. For inductance, the measurements agree with the designed value within 10\%, but tend to be lower.

The reduction of inductance in the measurements is likely due to imperfections in the coil fabrication, apart from systematic errors in the measurements. In visual inspection, the wire winding of the coils is sometimes random. Moreover, the outermost layer of the winding finishes in the middle as already mentioned, and that would reduce the effective number of the wire turns, $N$ in Eq.~(\ref{eq:coil_induct}); for example, the estimated inductance reduces from 8.9 to 8.6~mH by using 27 instead of 27.5 for the number of layers.

There are several potential issues with the existing actuator design. The current-carrying wire will expand or shrink in diameter under a magnetic field, and so the random winding would cause non-uniform stress to slip the wire loops, and that could become a noise source; the worst case would be breakage of the wire. Though there is no evidence for this at present, the coil fabrication should be improved in any redesign; using square wires may be helpful.

\subsection{Electromagnetic forces}
\begin{figure*}
\begin{center}
\includegraphics[width=16.8cm]{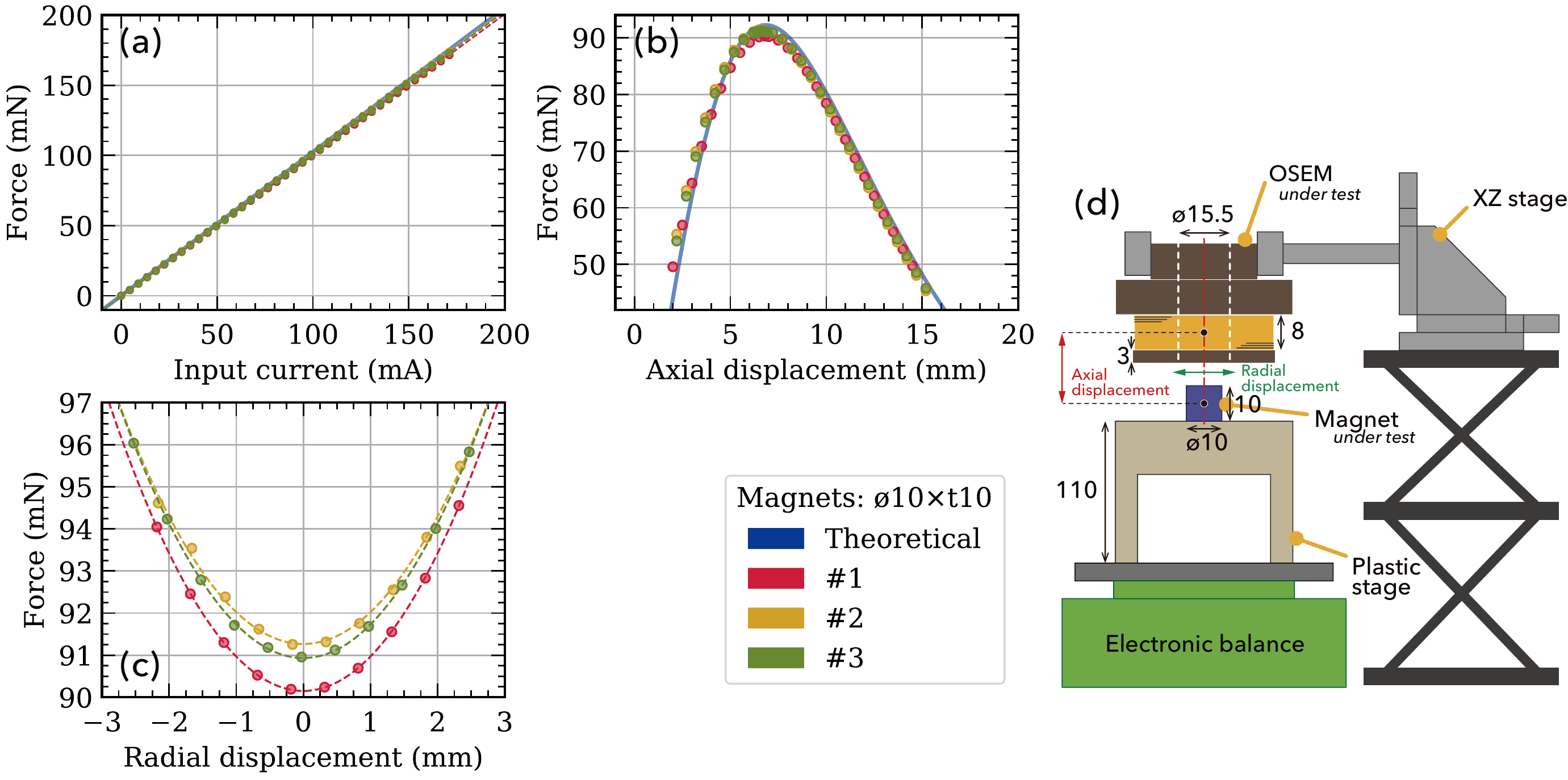}
\caption{Variations of the electromagnetic forces arising between a spare OSEM and three spare magnets in the same shape and material (SmCo); the forces with respect to (a) the current carried in the coil, (b) the displacement of the magnet in the axial direction, and (c) that in the radial. The filled circles show the measurements. In (a) and (c), fitted curves are drawn by the dashed lines; see also Table~\ref{tab:act_fit}. The solid blue curves in (a) and (b) are theoretical estimates from the design but not fitted. A schematic view of the measurement system is shown in (d).}
\label{fig:act_response}
\end{center}
\end{figure*}

We measured the electromagnetic forces with respect to the current carried in the coil, and to the displacement of the magnet in the axial and radial directions (Fig.~\ref{fig:act_response}).
For the measurements, we dedicated a spare OSEM left in our storage (but not included in Table~\ref{tab:responses}), together with three spare magnets of the same type as installed in KAGRA (see Section II).

Fig.~\ref{fig:act_response} (d) shows a schematic view of the measurement setup. The coil in the OSEM under test is located over the cylindrical permanent magnet under test. The coil is supported by a bar from the XZ translation stage with micrometers on a rigid structure. By adjusting the micrometers, the relative position of the coil and the magnet can be adjusted in the radial and axial directions. The ends of the coil wire are connected to a DC power supply, P4K-80L by Matsusada Precision.
The electromagnetic force arising on the coil and the magnet was measured by an electronic balance, BL-220H by Shimadzu. The magnet was raised on a handmade plastic stage to keep it well separated from the top surface of the balance. Without that distance, the magnet and the coil came too close to the electronic balance, and the magnetic field from the coil affected the reading of the balance, making the measurements of the electromagnetic force inaccurate.

A user-friendly actuator must have linearity from input to output.
In Fig.~\ref{fig:act_response} (a), the filled circles show the measurements of the electromagnetic forces arising from the current carried into the coil, while the dashed lines are the fitted curves. For every combination, the measured points can be fit by a linear function, and the coefficient is $\sim 1\,\mathrm{N/A}$; see Table~\ref{tab:act_fit}.
\begin{table}
\caption{\label{tab:act_fit}Fit parameters for each curve drawn in Fig.~\ref{fig:act_response} (a) and (c). Note that each set of measured points in (c) is offset along the horizontal axis to minimize the fitted quadratic curve at $x=0$.}
\begin{ruledtabular}
\begin{tabular}{lccc}
Coefficients &\#1 & \#2 & \#3 \\
\hline
\multicolumn{4}{c}{Fit curve: $y=a x +b$\quad ($a$: N/A, $b$: mN)}\\
Force/Current & 1.01, $-0.38$ & 1.02, $-0.37$ & 1.01, $-0.35$\\
\multicolumn{4}{c}{Fit curve: $y=a x^2 + b$\quad ($a$: $\mathrm{mN/mm^2}$, $b$: mN)}\\
Force/Radial disp. & 0.82, 90.1 & 0.78, 91.3 & 0.80, 90.9
\end{tabular}
\end{ruledtabular}
\end{table}%
As seen below, the force arising also depends on the displacement of the magnet relative to the coil in the axial and radial directions.
For each set of the measurements in Fig.~\ref{fig:act_response} (a), we positioned the magnet at a ``sweet spot'', $\sim 7$~mm axial position and $\sim 0$~mm radial position, so that the force derivative was zero in each degree of freedom as found in Fig.~\ref{fig:act_response} (b) and (c).
In the actual vibration isolation systems, the gap in each OSEM must be adjustable to achieve the sweet-spot operation to prevent unwanted changes of the coefficient during the operation.

The solid blue line in Fig.~\ref{fig:act_response} (a) shows an estimate from the design of the coil and a magnetic flux density at the surface on the end of the cylindrical magnet, $B_\mathrm{s}=409\,\mathrm{mT}$, which was measured with a teslameter, TM-801 by Kanetec. From this number, we can assume a point magnetic dipole moment equivalent to the cylindrical magnet; $m_z=0.57\,\mathrm{A\, m^2}$ being coaxial with the cylinder; see Appendix~\ref{Sec:force_by_coil}. By following the method written there, the theoretical estimate of the current-to-force coefficient is $\sim 1.025~\mathrm{N/A}$. As in Table~\ref{tab:act_fit}, the coefficients obtained from the measurements are very close to the theoretical estimate.

In order to find the sweet spot for each combination, we measured the electromagnetic forces while varying the distance between the coil and the magnet being on-axis in the axial direction; see Fig.~\ref{fig:act_response} (b). The current carried into the coil was fixed at about 90~mA; here, 1~V was applied to the coil with a power supply, and the resistance of the coil was measured as $11.1~\Omega$ for this OSEM under test.
In this figure, the horizontal axis shows the distance between the center points of the cylindrical magnet and the coil; for example, 0~mm means the magnet is located at the center of the coil. For every combination, the sweet spot, where the maximum force arises, is found at 6.6-6.7 mm.

The solid blue curve in Fig.~\ref{fig:act_response} (b) shows an estimate from the dipole equivalent to the cylindrical magnet discussed already. By a close look at the curve, the sweet spot can be found at $\sim 6.8\,\mathrm{mm}$ without relying on numerical search algorithms. As in Appendix~\ref{Sec:force_by_coil}, an analytic form of the curve can be derived, but it is difficult for us to find the sweet spot as an analytic solution. Despite a number of degenerated parameters in the analytic form of the curve, the measurements show a good agreement with the theoretical estimate. 

Similar measurements in the radial direction are shown in Fig.~\ref{fig:act_response} (c). We used a quadratic curve (empirically) for fitting every combination of the coil and the magnets, as the system is rotationally symmetric; see also Table~\ref{tab:act_fit}.
The cylindrical magnet has a diameter of 10~mm, while the central hole of the coil is 15.5~mm in inner diameter (Figs.~\ref{OSEM_all} and~\ref{OSEM_w_flag}), so the magnet can only move 2.75~mm at most in the radial direction. Unwanted decentering of the magnet can be easily discovered thanks to the narrow separation.

\section{Discussions}

\subsection{In comparison with the previous design}
This subsection compares the responses of the previous (ver.~1) and current design (ver.~2) of the optical sensors in the OSEMs.
As mentioned already, we widened the separation between the TX and the RX from 5~mm to 15~mm after the initial test run of KAGRA mainly to avoid the risk of mechanical interference with the sensor flag, and so the nominal gaps between the sensor flag and each of those units increased from 1.5~mm to 6.5~mm (Fig.~\ref{OSEM_w_flag}). 

As such, the modification was done without changing the OSEM body because we wanted to reuse as many parts as possible from the previous OSEMs (ver.~1). If we had a chance to design the OSEM body from scratch, we will revisit the whole mechanical design including the diameters of the holes for the TX and RX; for example, to fit the PD holder with the OSEM body, the outer diameter of the PD retainer has to be M11 threaded, which should have been avoided as it is not a normal size for manufacturers. 

\begin{figure}
\begin{center}
\includegraphics[width=8.5cm]{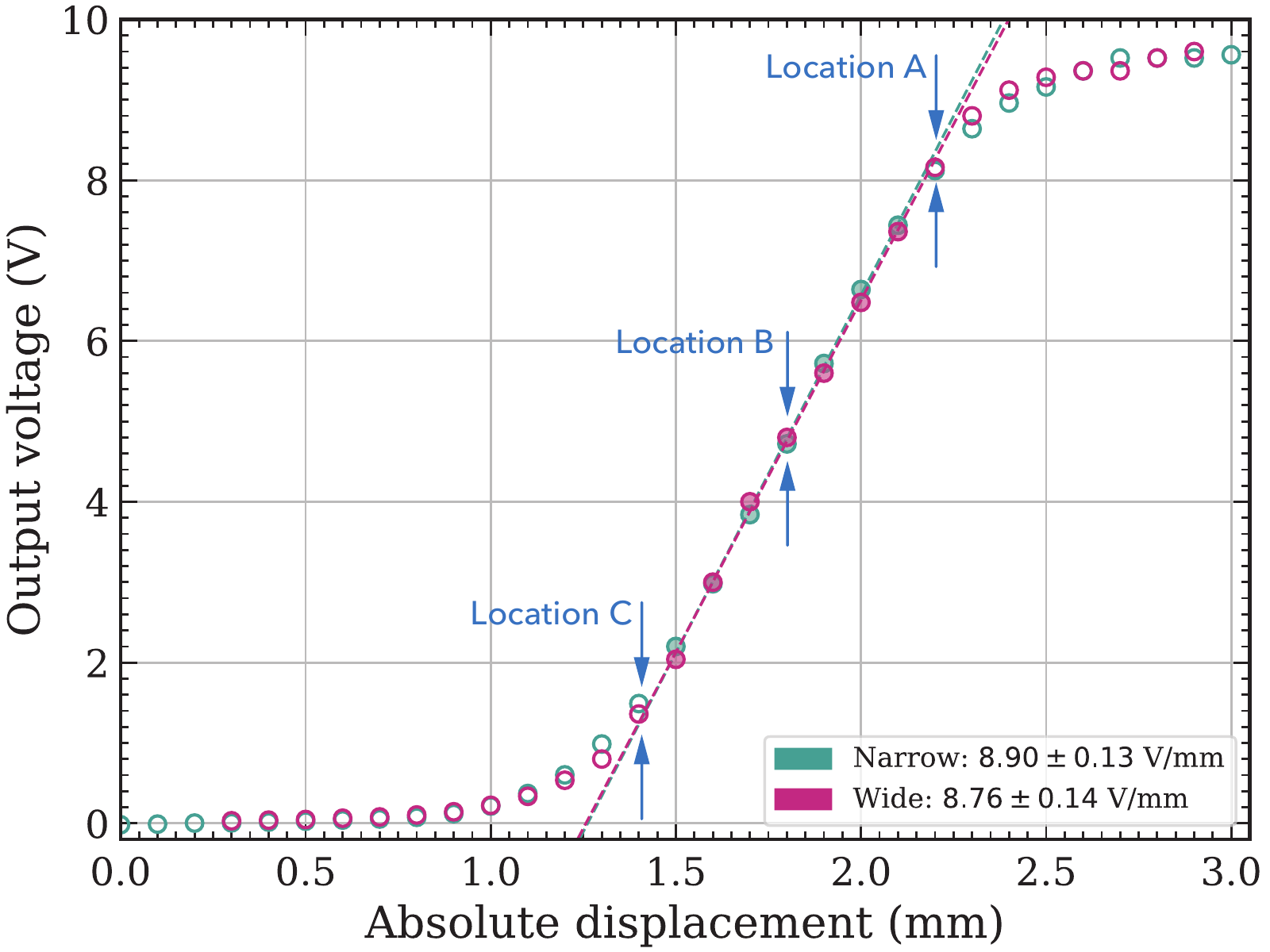}
\caption{Responses of the sensors of OSEM ver.~1 and 2; they are labeled ``narrow'' and ``wide'', respectively. Each response is represented by output voltage from a handmade test circuit with respect to the position of the sensor flag. The circles show measured data points. The filled circles are the data points used for linear fit, and the fit curves are drawn with dashed lines. Location A, B, and C are referred later.}
\label{comp_sens}
\end{center}
\end{figure}%
Before mass production of the current OSEMs (ver.~2), we prepared a prototype of ver.~2, and compared it with ver.~1.
Fig.~\ref{comp_sens} shows the responses of the prototype of ver.~2 and ver.~1 (labeled ``wide'' and ``narrow'', respectively) in the same manner as in Fig.~{\ref{SensCurves}} (a). For the measurements,
we used a handmade test circuit (single-ended signaling) rather than the driver circuit used in KAGRA.
In the test circuit, the LED driving current is adjustable with a variable resistor, and so we can vary the brightness of the LED. Note that the gain-setting resistors in the KAGRA driver circuits are non-variable and would need to be replaced to adjust the gain.
For comparing the responses of the two designs, 
we adjusted the LED driving current in each measurement so that the readout voltage became equal at the fully open location of the sensor flag (about 9.6~V at around 3~mm in Fig.~\ref{comp_sens}). The full-open readout from the ``wide'' prototype was half that of the ``narrow'' one without adjusting the LED driver current.

The dashed lines in Fig.~\ref{comp_sens} are linear fits to the data from the two measurements, and the slopes are $8.90\pm 0.13$ V/mm and $8.76\pm 0.14$ V/mm for the ``wide'' prototype and the ``narrow'' one, respectively. They are within the margin of error from each other.

The 50\% reduction in maximum output voltage is due to the increased separation between the TX and RX. In contrast, the responses do not vary. This invariance is partly due to the tight collimation of the light beam from the LED (discussed later). So far, we have driven the LEDs in the current OSEMs at the site without compensating for the reduction. If compensation is needed in the future, we will modify the driver circuits.

\subsection{Insensitivity to motion in other directions}
We had a concern that the wider separation of the TX and RX in the sensor would increase coupling from motion in other degrees of freedom. Thus, we evaluated the coupling coefficients with a prototype of the current OSEM (ver.~2) sensor before mass production to confirm negligible coupling.

\begin{figure}
\begin{center}
\includegraphics[width=8.5cm]{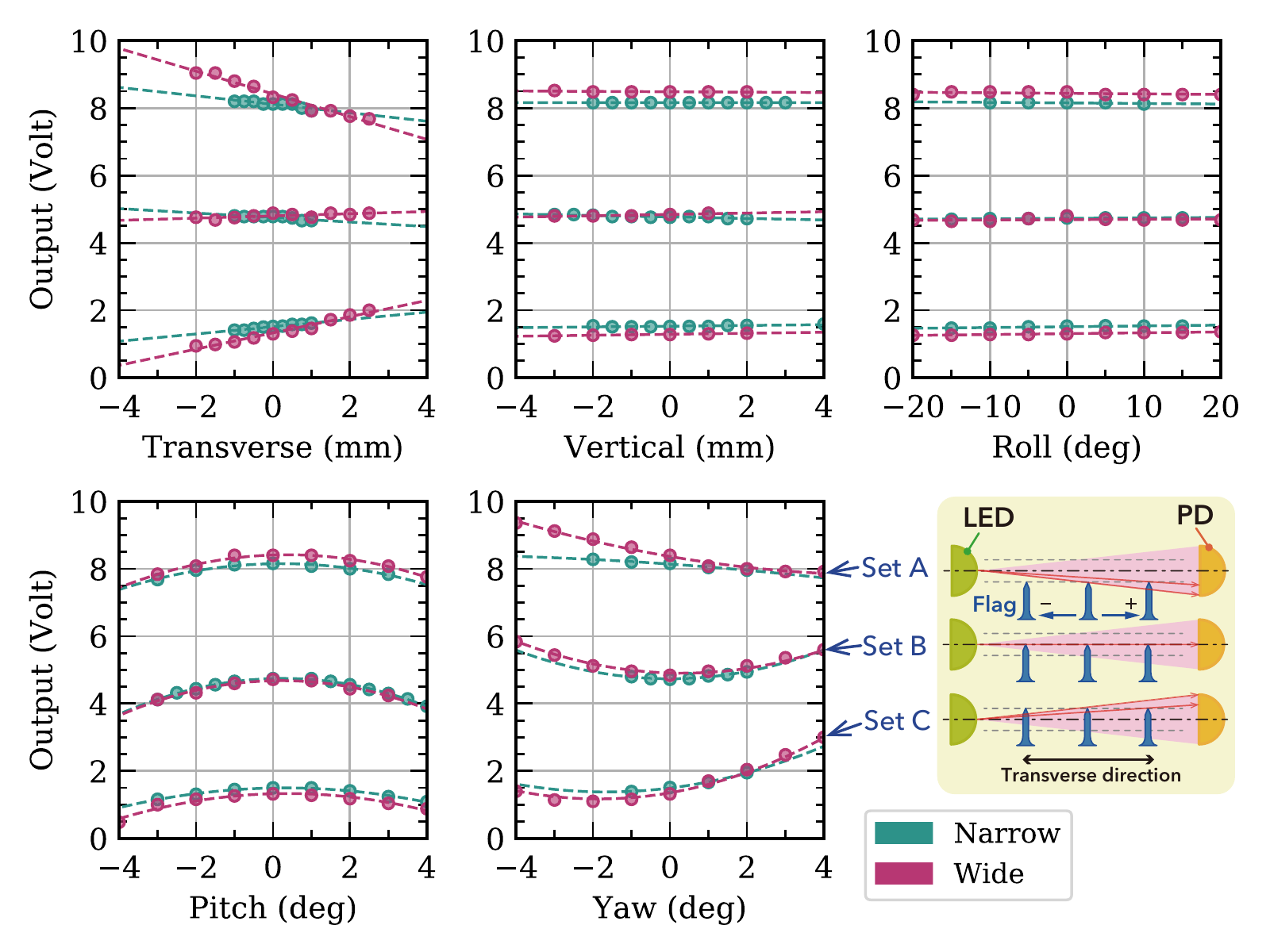}
\caption{Responses of the sensors of OSEM ver.~1 and 2 with respect to the other degrees of freedom; they are labeled ``narrow'' and ``wide'', respectively, in the same manner as Fig.~\ref{comp_sens}. In each subfigure, the data points are categorized into three sets denoted set A, B, and C, collected when the longitudinal position of the tip of the sensor flag was fixed at location A, B, or C in Fig.~\ref{comp_sens}, respectively. The fit parameters are summarized in Table~\ref{tab:fitting2}. An exaggerated schematic view that would explain the response to the transverse motion of the sensor flag is shown at the right bottom.}
\label{couple_dofs}
\end{center}
\end{figure}

Fig.~\ref{couple_dofs} shows the measured variation of the output voltages of the sensor ``wide'' in response to motions in the other degrees of freedom; their fitted curves are also drawn. Estimates of the coupling coefficients from the fitted curves are summarized in Table~\ref{tab:fitting2}. As in the case of Fig.~\ref{comp_sens}, we measured them with one of the previous ``narrow'' OSEMs as well, and the measurements are also shown in Fig.~\ref{couple_dofs}.
For the measurements, the handmade test circuit introduced in the previous subsection was used, and so the LED driver current was adjusted when measuring the ``wide'' curves as in the previous subsection.
In addition,  the measurements shown in Fig.~\ref{couple_dofs} can be directly compared with those in Fig.~\ref{comp_sens}.
\begin{table*}
\caption{\label{tab:fitting2}Fit parameters for each curve drawn in Fig.~\ref{couple_dofs}. The responses to transverse, vertical, and roll motions are fitted to a linear function ($y=a x +\mathrm{Const.}$), and estimates of the slopes are shown. For pitch and yaw, quadratic curves ($y=ax^2+bx+\mathrm{Const.}$) are chosen for fit; note that the slopes at $x=0$ correspond to $b$ for such quadratic curves. }
\begin{ruledtabular}
\begin{tabular}{lcccc}
Degrees of freedom & &Set A & Set B & Set C\\ \hline
\multicolumn{5}{c}{Fit curve: $y=a x +\mathrm{Const.}$\quad ($a$: V/mm or V/deg)}\\
Transverse & Narrow & -0.125(23)& -0.066(15) & 0.108(6) \\
      & Wide & -0.337(20) & 0.032(11) & 0.240(15)\\
Vertical & Narrow &  0.0000(0) & -0.0222(36) & 0.0111(31) \\
      & Wide & -0.0057(33) & 0.0200(61) & 0.0149(14) \\
Roll & Narrow &  -0.0016(9) & 0.0013(3) & 0.0024(6) \\
      & Wide &  -0.0016(6) & 0.0010(9) & 0.0026(2) \\
\multicolumn{5}{c}{Fit curve: $y=a x^2 +bx+\mathrm{Const.}$\quad ($a$: V/deg$^2$, $b$: V/deg)}\\
Pitch & Narrow & -0.0433(38), 0.0186(66) & -0.0589(10), 0.0292(21) & -0.0314(16), 0.0200(36)\\
      & Wide & -0.0510(36), 0.0386(81)& -0.0586(63), 0.024(11)& -0.0389(43), 0.0297(97)\\
Yaw & Narrow & -0.0057(54), -0.0800(64)	 & 0.0533(87), 0.0010(115) & 0.042(12), 0.140(16)\\
      & Wide & 0.0171(40), -0.1947(90) & 0.0507(23), -0.0327(65) & 0.0529(30), 0.1967(83) \\
\end{tabular}
\end{ruledtabular}
\end{table*}%

\begin{figure}
\begin{center}
\includegraphics[width=8.5cm]{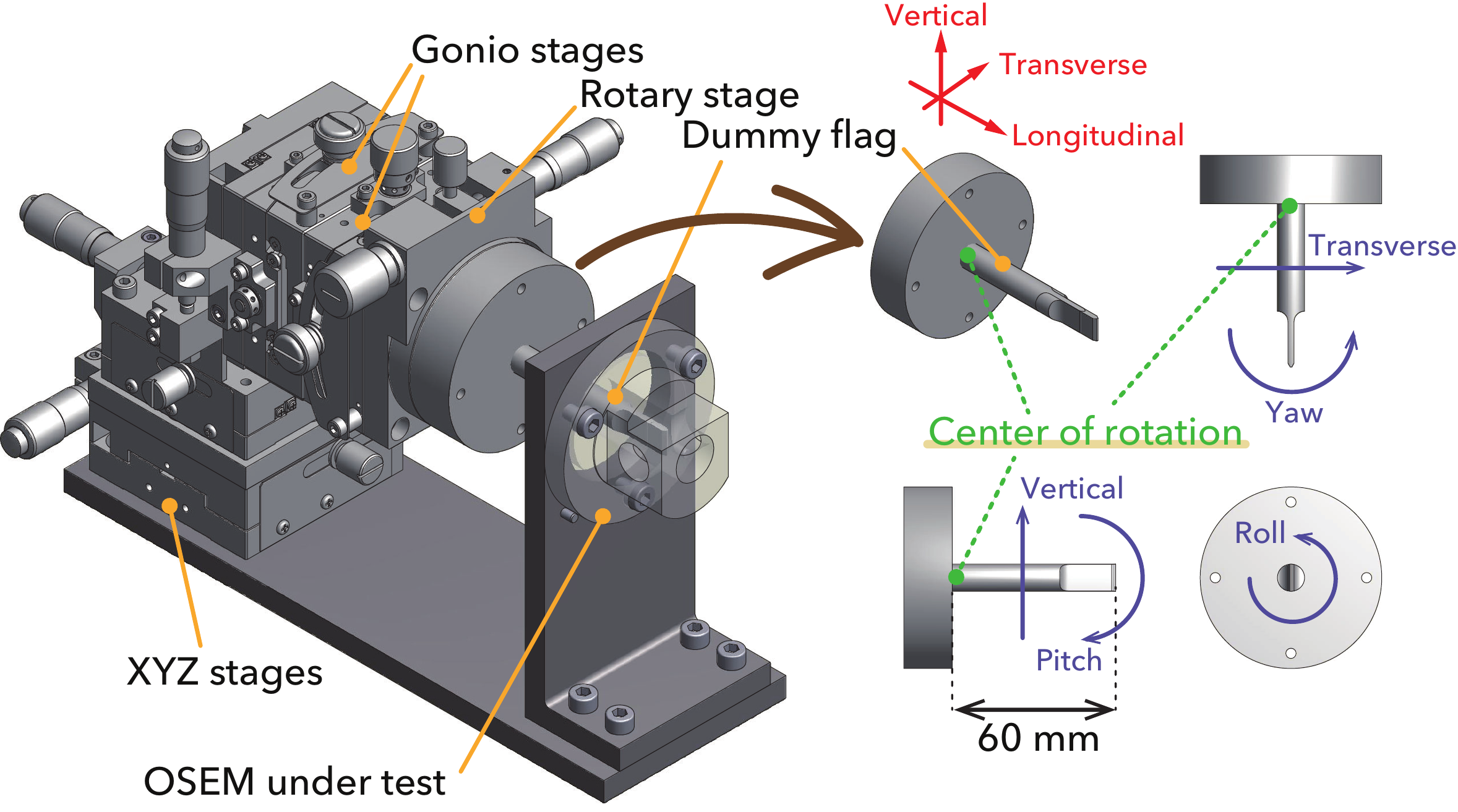}
\caption{Schematic view of a testbench for measuring coupling coefficients to the longitudinal sensor from the other degrees of freedom. A dummy sensor flag is attached onto the stack of stages movable in six degrees of freedom. The center of rotational motions is located at the base of the sensor flag in the design (60~mm below the tip of the flag). This setup is not clean-room compatible unlike the one shown in Fig.~\ref{SensCurves}, which is simpler and used during the actual mass production of the OSEMs.}
\label{normal_testbench}
\end{center}
\end{figure}

For the measurements, we prepared a dedicated testbench (Fig.~\ref{normal_testbench}), with which we can move the dummy sensor flag in the transverse, vertical, roll, pitch, or yaw directions, as well as the (nominal) longitudinal direction; see the figure for the definition of the directions.

The measurements were done while fixing the flag tip at three different longitudinal locations, A, B, and C in Fig.~\ref{comp_sens}. 
The data set collected at each of the locations is indicated with the corresponding letter, A, B, or C in Fig.~\ref{couple_dofs}. B is the nominal location of the sensor. A and C correspond to the upper and lower outer edges of the linear range, respectively.

As summarized in Table~\ref{tab:fitting2}, the responses to transverse, vertical, and roll motions were evaluated with linear fits, while quadratic curves were chosen for pitch and yaw; these were judged by eye to be the minimum degrees of polynomial to give a good fit.

For the vertical motion, every data set shows almost no slope. It is at most 0.02~V/mm in absolute value, which is negligible in comparison with the response in the longitudinal direction of $\sim$~8~V/mm. In other words, the coupling coefficient from vertical to longitudinal can be calculated as $0.02/8 \sim 2\times 10^{-3}\,\mathrm{mm/mm}$. The flag tip is 9~mm in vertical width, which is larger than the vertical width of the intensity profile of the light beam (about 7~mm; see (e) or (f) in Fig.~\ref{LightProfiles}), so it is unsurprising that vertical shift of the sensor flag does not change the shading of the light beam.

The same explanation can be applied to the roll motion. Even a 20-deg rolling of the sensor flag would change the vertical tip width seen from the RX by $\cos 20^\circ$, which corresponds to 8.5 mm, and is still larger than 7 mm. For the pitch motion, whether the flag rotates in positive or negative, the tip shades the light beam in the same manner, and so the responses show even-function behaviors like quadratic. In the case of the quadratic form like $y=ax^2 +bx +c$, the slope at $x=0$ is $b$. Thus, the pitch response around the nominal setup is about or less than 0.03~V/deg according to Table~\ref{tab:fitting2}, and the coupling coefficient from pitch to longitudinal can be calculated as $\sim 3\times 10^{-3}\,\mathrm{mm/deg}$ or $0.2\,\mathrm{mm/rad}$. Note that the possible rotation in pitch or yaw would be less than 0.2~rad due to mechanical limits imposed by the OSEM central hole.

A similar explanation can be applied to yaw motion, but the minimum point of each of the quadratic curves slightly varies with respect to the longitudinal locations of the sensor flag tip (A, B, and C). 
The variation can be explained as a combination of the quadratic and slope responses to the transverse motion of the flag tip (due to the yawing) discussed next.
We are still not sure of the reason for the difference between the ``narrow'' and ``wide'' responses at location A.

For the transverse motion, the estimated slopes vary greatly depending on the longitudinal locations of the flag tip. The slope at A is 10 times larger (in absolute value) than that of B for the ``wide'' design.
By comparing the behaviors of the ``wide'' and ``narrow'' OSEMs at A and B, the slopes are two or more times larger (in absolute value) for the ``wide'' than that for the ``narrow''.
The slope behaviors can be explained by a simple model assuming that the light beam from the LED to the PD is slightly expanding, which is depicted in an exaggerated schematic view at the right bottom in Fig.~\ref{couple_dofs}. The assumption is supported by the fact that the ``wide'' sensor has a smaller (about a half) photocurrent than the ``narrow'' one. In this assumption, the light power reached the PD is decreasing when the sensor flag, which longitudinal location is at A, is approaching to the PD, and the opposite happens at C, while at B the transverse motion of the flag does not change the light power at the PD. We could model the real beam profile as a combination of the collimated and expanding components, and so the ``narrow'' sensor, the PD of which is located in the collimated region, would have shallower slopes; see also Fig.~\ref{LightProfiles} and the related discussion in the following subsection about the beam profiles.
So far, the couplings are so small that they are not practically problematic for KAGRA, but further investigation could be required in the future.

\subsection{Lens and slit in the TX}
\begin{figure}
\begin{center}
\includegraphics[width=8cm]{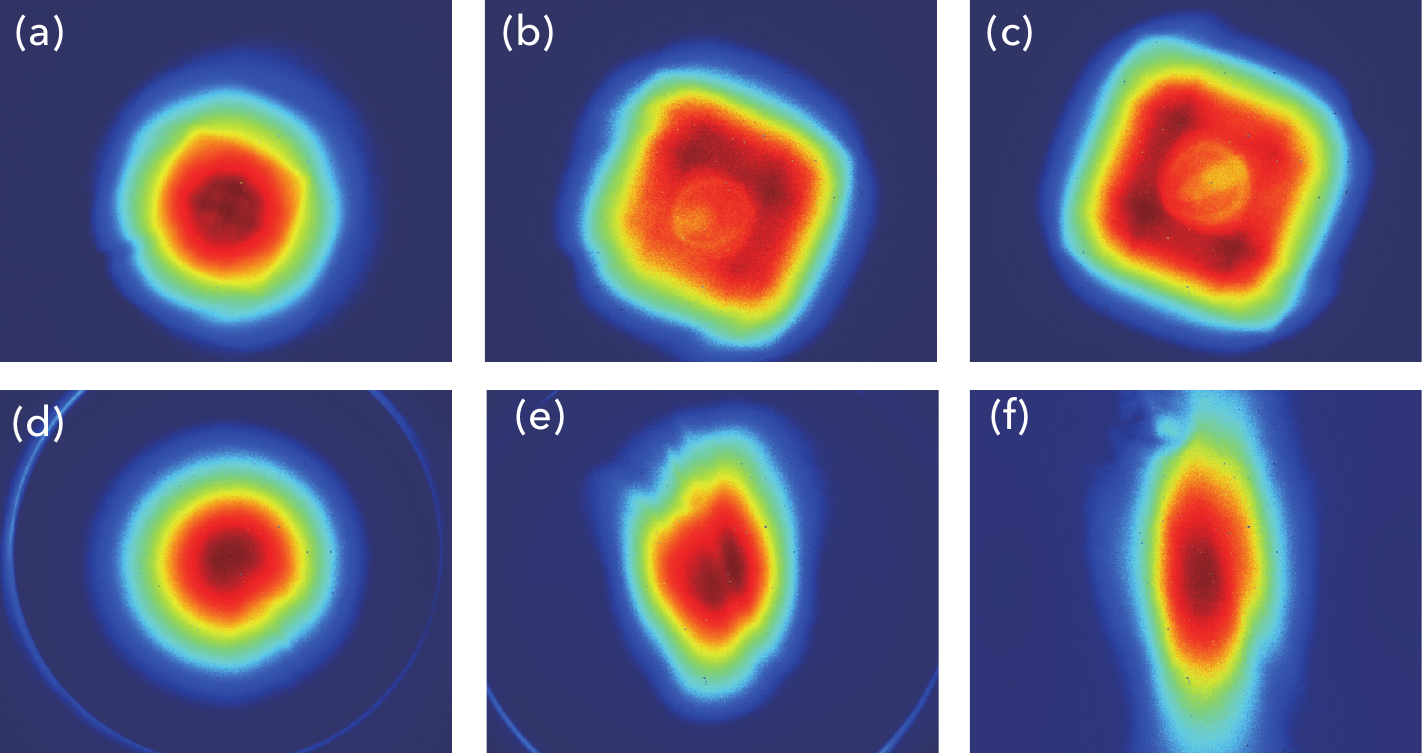}
\caption{Emission profiles of TX prototypes in various conditions; the contours represent irradiance of the light. (a) measured at 12~mm distance from the LED without the lens or the slit lid; (b) at 22~mm; (c) at 26~mm; (d) at 26~mm with the lens in front of the LED; (e) with the lens and the slit lid; (f) at 34.5~mm with the lens and the slit lid. Each subfigure corresponds to 6.8~mm $\times$ 5.4~mm. Note that the nominal separation between the LED and the sensitive area of the PD is 28.7~mm in the design.
}
\label{LightProfiles}
\end{center}
\end{figure}

We need the lens in the TX of the sensor to make the emission profile of light from the LED smooth and uniform, as in the LIGO OSEMs~\cite{Aston:2012,Carbone:2012}. A non-uniform profile will distort the linear response of the sensor. We measured the emission profiles of the LEDs in the TX prototypes under various conditions (Fig.~\ref{LightProfiles}).
For the measurements, we used a beam profiler, CinCam CMOS-1202 by Cinogy, which has an active area of 6.8~mm $\times$ 5.4~mm.

In Fig.~\ref{LightProfiles}, (a), (b), and (c) show the emission profiles measured at distance of 12~mm, 22~mm, and 26~mm, respectively, from the LED; (a) and (b) were obtained from the identical LED, but (c) was not. For the measurements, we detached the built-in lens and the slit lid from the TX (see Figs.~\ref{OSEM_sensor} and \ref{OSEM_w_flag}). The spatial profile departs from a uniform shape to a spotted square-like shape with increasing distance. Looking closely at the center of (a), one can find a seed of such a spotted square pattern. The spotted square indicates an image of the emitting chip and the shadows of wires to it inside the LED package.
In the current design,
the nominal separation from the LED to the sensitive area of the PD is 28.7~mm. If we let the profile propagate, the non-uniform light illuminates the PD, and distorts the linearity of the sensor response.

The profile in (c) changed to (d) when we attached the lens at the nominal separation from the LED; we used an identical LED for measuring the profiles from (c) to (f). In (d), the profile becomes smoother and better concentrated than that in (c). The drawback is a halo surrounding the concentrated area. We have not yet understood that; probably it is due to spherical aberration or a diffraction pattern. So far the halo has not noticeably distorted the linear response of the sensors (Fig.~\ref{SensCurves}), but may become an issue in the future. For example, the halo is stray light for the sensor, and might cause increased noise.

The profile in (d) changed to (e) when the slit lid and the lens were both attached. We expected the slit aperture to limit the light profile illuminating the PD, and cut out unnecessary light. The drawback of the slit aperture is apparent in (e); the light profile is unexpectedly distorted. So far, it has not distorted the linear response of the sensors (see Fig.~\ref{SensCurves}), so we are satisfied with it for now.

A simple circular aperture might be even better to avoid the distortion of the profile, but we need careful evaluation from several perspectives for new apertures. The profile became smoother at a location further away as shown in (f), where the distance from the LED was 34.5~mm. The profile expanded in not only the vertical but also the horizontal direction with a broad tail, and the halo went out of the measurement area. As described above, the PD is located at 28.7~mm, and so it is illuminated with a profile in between (e) and (f).

\subsection{Eddy current loss}
Eddy current loss in the actuator body can be a severe noise source for the suspensions~\cite{Agatsuma:2010}. As was mentioned above, the OSEM body is made of carbon-loaded PEEK, Ketron CA30. Although this type of material usually shows very high electrical resistivity $\rho\sim 10^3\,\mathrm{\Omega\cdot m}$, our measurements show lower values.
In LIGO, they also found this material was too conductive~\cite{Barton:2010B}, and changed to a more suitable one, ESd 480 by Semitron, which has a clear description of the lower limit of the resistivity in the specification. Thus, let us quantify the energy loss and evaluate the detrimental impact.

\begin{figure}
\begin{center}
\includegraphics[width=8.6cm]{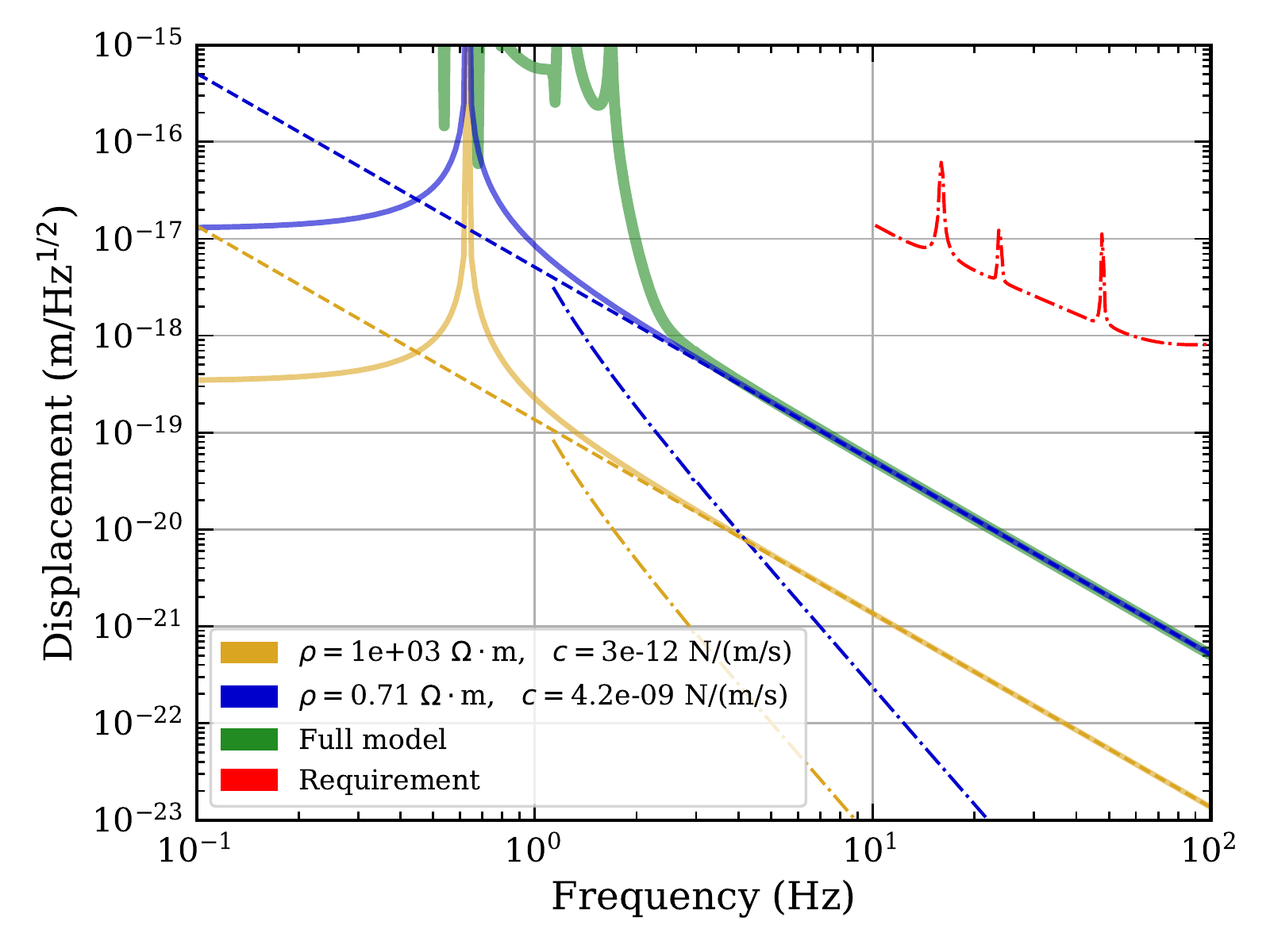}
\caption{One-sided amplitude spectral density of the calculated thermal noise contribution due to eddy current loss in the OSEM body for the selected electrical resistivity $\rho$ and the corresponding damping coefficient $c$. The solid blue and yellow lines show the contribution to the motion of the intermediate mass (IM) in the beam splitter (BS) suspension of KAGRA using a simple model; the dashed lines show their asymptotes in the high frequency region. The solid thick green line shows the contribution calculated with a full suspension simulator for the blue line case. The IM displacement noises are converted to the BS motion and drawn with the dash-dotted lines in the corresponding colors. For comparison, the requirement for the BS motion to achieve the KAGRA sensitivity, which is the same one in Fig.~6~(b) of Ref.~\cite{Michimura:2017}, is also shown with the dash-dotted red line.}
\label{fig:ThermalNoise}
\end{center}
\end{figure}%

Already in the early days of the GW detectors~\cite{Miyoki:1997}, one could not neglect the eddy current loss of coil-magnet actuators in vibration-isolation systems for optics. The loss in the actuator will cause unwanted viscous damping, and so reduce the vibration-isolation ratio. In addition, the loss will introduce thermal noise into the system. The noise process has been usually explained by the fluctuation-dissipation theorem~\cite{Saulson:1990,Gillespie:1994}.
Hereafter we will take the dynamic model in Eqs.~(\ref{eqs:two-body_mech}) as an example for discussing the eddy current loss.

Like the other resonant systems, the loss in a pendulum can be characterized in terms of $Q$-factor at the resonance or loss angle $\phi$~\cite{Cagnoli:1998,Frasca:1999,Barton:2000,Strain:2005}:
\begin{align}
Q_1&=\frac{m_1\omega_1}{c}=\frac{1}{2\xi_1},\label{eq:Q-factor}\\
\phi&=\frac{\omega}{Q_1\omega_1},
\end{align}
where the notation follows Eqs.~(\ref{eqs:two-body_mech}). Usually, we aim to retain $Q\sim 10^5$ or more for our pendulums when freely swinging. By assuming $m_1\sim 1\,\mathrm{kg}$ and $\omega_1\sim 2\pi\times 1\,\mathrm{Hz}$, the net amount of the viscous damping coefficient $c$ should be $\sim 6\times 10^{-5}\,\mathrm{N/(m/s)}$ or less. One of the components of $c$ comes from the eddy current loss as summarized in Appendix~\ref{Sec:eddy_current_loss}, and the theoretical estimate can be calculated with Eq.~(\ref{eq:viscous_eddy}).

For the calculation, let us assume the magnetic dipole $m_z$ to be the value we discussed in the previous section, and the electric resistivity of the bobbin material to be $\rho\sim 10^3\,\mathrm{\Omega\cdot m}$ from the specification. In Eq.~(\ref{eq:viscous_eddy}), taking the integral limits as $(r_1,r_2) = (7.75, 9.0)\,\mathrm{mm}$ and $(z_1,z_2)=(-0.3, 41.7)\,\mathrm{mm}$, we obtain $c\sim 3\times 10^{-12}\,\mathrm{N/(m/s)}$, which is well below the target value. In fact, the integration range is not so obvious especially for $r_2$, as the actual bobbin has a complicated structure. The integral, however, converges rapidly as increasing $r_2$, and the resultant $c$ will be at most $\sim 8\times 10^{-12}\,\mathrm{N/(m/s)}$; the order of magnitude does not change so much.

The OSEM body had almost no conductivity as expected, but we found out that low resistances were sometimes observed between small screw holes on the OSEM body, while relatively wider surfaces were well insulated. We do not fully understand the origin of the issue yet.

Suppose $\sim 90\,\Omega$ was measured between such two screw holes at 1~cm distance. Estimating an effective cross section of the current path is difficult but let us assume a $\sim 1$~cm diameter circle. This leads $\rho\sim 0.7\,\Omega\cdot\mathrm{m}$. In fact, the specification for the material just mentions the upper limit of the resistivity such as $<\sim 10^5\,\Omega\cdot\mathrm{m}$, but not for the lower limit, and so literally the measurement here satisfies the specification. As in Eq.~(\ref{eq:viscous_eddy}), however, $c$ is proportional to $\rho^{-1}$, and so it increases to $c\sim 10^{-9}\,\mathrm{N/(m/s)}$. From the viewpoint of thermal noise due to the eddy current loss, the large $c$ might be problematic as discussed below.

In the dynamic model in Eqs.~(\ref{eqs:two-body_mech}), the thermal noise contribution at the suspended mass in Fig.~\ref{concept0} (a) can be calculated as~\cite{Saulson:1990}:
\begin{equation}
S_x^2 = \frac{4k_\mathrm{B}T}{\omega^2}\,\mathrm{Re}\left[Z^{-1}\right],\label{eq:th_noise}
\end{equation}
where $k_\mathrm{B}$ is the Boltzmann constant, $T$ is the temperature of the system, and $Z$ is the mechanical impedance of the suspended mass. Note that $S_x$ is a one-sided amplitude spectrum density of the suspended-mass fluctuations due to thermal noise. To derive $Z$, assuming a fluctuation force $F_1$ on the suspended mass, we can rewrite Eqs.~(\ref{eqs:two-body_mech}) as
\begin{subequations}
\begin{align}
&m_1 \ddot{x}_1 + c(\dot{x}_1-\dot{x}_2)+k_1x_1=F_1,\\
&m_2 \ddot{x}_2 + c(\dot{x}_2-\dot{x}_1)+k_2x_2=0.
\end{align}\label{eqs:two-body_mech_thermal}
\end{subequations}
In the same way as from Eqs.~(\ref{eqs:two-body_mech}) to (\ref{eq:ratio}), we can obtain
\begin{equation}
\frac{\tilde{x}_1}{\tilde{F}_1}=\frac{1}{k_1}\frac{1+2i\xi_2\frac{\omega}{\omega_2}-\frac{\omega^2}{\omega_2^2}}{(1+2i\xi_1\frac{\omega}{\omega_1}-\frac{\omega^2}{\omega_1^2})(1+2i\xi_2\frac{\omega}{\omega_2}-\frac{\omega^2}{\omega_2^2})+4\xi_1\xi_2\frac{\omega^2}{\omega_2\omega_1}}\label{eq:xF_ratio}
\end{equation}
in the frequency domain. From this relation, the mechanical impedance is derived as $Z=\tilde{F}_1/(i\omega \tilde{x}_1)$. 

In Fig.~\ref{fig:ThermalNoise}, the solid blue and yellow curves show $S_x$ for two different $\rho$; we assume the temperature $T=300\,\mathrm{K}$, the masses $(m_1, m_2) = (41.7, 15.8)\,\mathrm{kg}$, and the resonant frequencies $(f_1, f_2)=(0.7, 1.2)\,\mathrm{Hz}$, respectively. These input parameters are of the intermediate mass (IM) and the recoil mass for the intermediate mass (IRM), respectively, for the beam splitter (BS) suspension in the KAGRA interferometer. The noise spectra are proportional to $\sim f^{-2}$ above $f_1$. The dashed lines show the asymptotes $S_x\sim (4k_\mathrm{B}T\omega_1/(m_1 Q_1 \omega^4))^{1/2}$, where we defined $Q_1$ in Eq.~(\ref{eq:Q-factor}).
For reference, we calculated the corresponding thermal noise of the intermediate mass with full suspension-model simulation with SUMCON~\footnote{\url{https://gwdoc.icrr.u-tokyo.ac.jp/cgi-bin/DocDB/ShowDocument?docid=3729}} for the case of the blue curve, and that is shown with the solid thick green line; they are consistent with each other in the high frequency region.

The BS is disturbed by the thermal noise from the IM stage. Considering the BS is suspended from the IM with a pendulum having a resonant frequency of 0.7~Hz, we can calculate the thermal noise contribution at the BS stage by taking care of the mechanical filtering effect, and they are shown with dash-dotted lines in Fig.~\ref{fig:ThermalNoise}. They are well smaller than the requirement for the BS motion shown with the dash-dotted red line, which is the same one in Fig.~6~(b) of Ref.~\cite{Michimura:2017}. A similar discussion stands for the other room-temperature suspensions in the KAGRA interferometer. The thermal noise due to eddy current loss in the OSEMs installed at the penultimate stage (or IM) will not significantly affect fluctuations of the mirror at the lowest stage. In other words, we do not need to change the material of the OSEM body if the true value of $\rho$ is not far from our assumption.

As already mentioned, complete OSEMs are only installed at the penultimate stage of the pendulums at KAGRA today. However we also need to consider the solenoid actuators attached to the recoil mass for the (room-temperature) mirrors, which are made of the same material; this point will be discussed in future publications.

\section{Conclusions}
In this paper, we have described a compact module integrating an optical sensor and an electromagnetic actuator (OSEM). The module is especially for vibration-isolation systems in the GW detector KAGRA, where 42 are in operation without apparent issues as of 2020. The sensor part was modified after the initial test run of KAGRA in 2016, according to the lessons we learned during the run. The modification is to widen the spacing of the sensor slot from 5~mm to 15~mm to avoid the risk of mechanical interference with the sensor flag.
In order to investigate the effect on the module performance due to this modification, we characterized the modified module. The sensor noise is about 0.5~nm/Hz$^{1/2}$ at 1~Hz, and 0.1~nm/Hz$^{1/2}$ at 10~Hz. The actuation coefficient is 1~N/A. 
The responses of the sensor and actuator in the extraneous degrees of freedom are negligible.
Some potential concerns in the design including eddy current loss of the module body are also discussed.  

\begin{acknowledgments}
We thank Takanori Sekiguchi, Joris van Heijningen, and Daisuke Tatsumi for prototyping our OSEMs in the early days. We also thank the Advanced Technology Center (ATC) of NAOJ for giving priority to the KAGRA project, and supporting us with maintaining experimental rooms, clean environment, and precision cleaning systems, and for mechanical engineering and advisement.
This work was supported by MEXT, JSPS Leading-edge Research Infrastructure Program, JSPS Grant-in-Aid for Specially Promoted Research 26000005, JSPS Grant-in-Aid for Scientific Research on Innovative Areas 2905: JP17H06358, JP17H06361 and JP17H06364, JSPS Core-to-Core Program A. Advanced Research Networks, JSPS Grant-in-Aid for Scientific Research (S) 17H06133, the joint research program of the Institute for Cosmic Ray Research, University of Tokyo, National Research Foundation (NRF) and Computing Infrastructure Project of KISTI-GSDC in Korea, Academia Sinica (AS), AS Grid Center (ASGC) and the Ministry of Science and Technology (MoST) in Taiwan under grants including AS-CDA-105-M06, the LIGO project, and the Virgo project.
\end{acknowledgments}

\section*{Data Availability}
The data that support the findings of this study are available from the corresponding author upon reasonable request.

\appendix

\section{Inductance of a multilayer solenoid}
\label{Sec:calc_induct}
To calculate the inductance of a coil, let us assume each turn of the coil to be an individual closed loop of conductor. The coil can be approximated as an accumulation of such loops.

In practice, self-inductance $L$ of the coil can be estimated by a summation:
\begin{equation}
L = \sum_{i=1}^{N}L_i+\sum_{\substack{i,j=1\\ (i\neq j)}}^{N}M_{ij}\label{eq:coil_induct},
\end{equation}
where $N$ is the total number of the loops, $L_i$ is the self-inductance of each loop for $i=1\dots N$, and $M_{ij}$ is the mutual inductance between the loops $i$ and $j$. This is a summation of the self-inductance of every loop and the mutual inductance of all the possible combinations among the loops.

The mutual inductance is calculated by Neumann's formula.
According to the formula, mutual inductance of a pair of loops in a space with magnetic permeability $\mu$ is given by
\begin{equation}
M_{ij} = \frac{\mu}{4\pi}\oint_{C_i}\oint_{C_j}\frac{\cos\theta ds_ids_j}{l}\label{eq:Neumann}
\end{equation}
where $ds_i$ and $ds_j$ are line elements along the closed loops $C_i$ and $C_j$, respectively; $l$ and $\theta$ are the distance and angle between $ds_i$ and $ds_j$, respectively. Note that $M_{ij}=M_{ji}$. 

In the case of the multilayer solenoid for our purpose, every loop is circular and aligned along the same axis, and so Eq.~(\ref{eq:Neumann}) reduces to~\cite{Maxwell:1954}
\begin{equation}
M_{ij} = \mu\sqrt{r_ir_j}\left[
\left(\frac{2}{k}-k\right)K(k)-\frac{2}{k}E(k)
\right],~\label{eq:Mij}
\end{equation}
where $r_i$ and $r_j$ are the radii of the loops $i$ and $j$, respectively, the modulus $k\equiv\sqrt{4r_ir_j/((r_i+r_j)^2+d^2)}$, and $d$ is the center-to-center distance of the loops. $K(k)$ and $E(k)$ are the complete elliptic integrals of the first and second kind in the following forms:
\begin{subequations}
\begin{eqnarray}
K(k)= \int_0^{\pi/2}(1-k^2\sin^2\varphi)^{-1/2}\,d\varphi,~\label{eq:El1}\\
E(k)= \int_0^{\pi/2}(1-k^2\sin^2\varphi)^{1/2}\, d\varphi.
\end{eqnarray}
\end{subequations}

In the same manner, the self-inductance of each loop, $L_i$, can be calculated by Eq.~(\ref{eq:Mij}) with $i=j$, but $k$ must be changed by redefining $d$ as the geometric mean distance
$d=r_\mathrm{w}\exp(-1/4)$, where $r_\mathrm{w}$ is a cross-sectional radius of the conductor in the electric wire.

We wrote a simple Python script to compute $L$ in Eq.~(\ref{eq:coil_induct}).
For computation of $M_{ij}$, Eq.~(\ref{eq:Mij}) is transformed to
\begin{equation}
M_{ij} = \mu\sqrt{r_ir_j}\frac{2}{\sqrt{k_1}}\left(K(k_1)-E(k_1)\right),
\end{equation}
where the new modulus $k_1\equiv (1-k')/(1+k')$ and $k'\equiv\sqrt{1-k^2}$, otherwise the integrand in Eq.~(\ref{eq:El1}) would not converge when $d\rightarrow 0$. Our script computed $K(k_1)-E(k_1)$ by the arithmetic-geometric mean for quick convergence~\footnote{Robert Weaver's website, \url{http://electronbunker.ca/eb/CalcMethods.html}}.

\section{Electromagnetic force by a solenoid actuator}
\label{Sec:force_by_coil}
In this section, we summarize useful formulae for designing a solenoid actuator, especially for estimating the electromagnetic forces arising between a multilayer solenoid and a cylindrical magnet. 

Suppose a point magnetic dipole moment $\vec{m}$ at rest is in an external field of magnetic flux density $\vec{B}_\mathrm{ext}$. Then the field exerts an electromagnetic force $\vec{F}$ on the dipole moment as~\cite{Greene:1971,Boyer:1988,Vaidman:1990,Hnizdo:1992,Jackson:1998}
\begin{align}
\vec{F}&=\nabla(\vec{m}\cdot\vec{B}_\mathrm{ext})-\frac{1}{c_0^2}\frac{d}{dt}(\vec{m}\times\vec{E}_\mathrm{ext})\\
&=(\vec{m}\cdot\nabla)\vec{B}_\mathrm{ext},
\end{align}
where $c_0$ is the speed of light in vacuum, and $\vec{E}_\mathrm{ext}$ is an external electric field that can be taken to be zero for our application.

Let us take the cylindrical coordinates $(r,\theta,z)$ with basis vectors $\{\bm{e}_r, \bm{e}_\theta,\bm{e}_z\}$, as the actuator, consisting of the coil and magnet, is cylindrically symmetric in the nominal setup; let us align the $z$-axis onto the common axis of the coil and the magnet.
Suppose the magnetic dipole is a constant: $\vec{m}=m_z\bm{e}_z$.
Then the force can be calculated as $\vec{F}=\nabla(\vec{m}\cdot\vec{B}_\mathrm{ext}) = \nabla (m_zB_z)$, and so
\begin{equation}
\vec{F}=m_z\left(\bm{e}_r\frac{\partial B_z}{\partial r}+\bm{e}_\theta\frac{1}{r}\frac{\partial B_z}{\partial\theta}+\bm{e}_z\frac{\partial B_z}{\partial z}\right),
\end{equation}
where $B_z$ is the $z$-component of $\vec{B}_\mathrm{ext}$.
Our main concern in this system is the $z$-component of $\vec{F}$. Specifically~\footnote{Note that the other components of $\vec{F}$ are supposed to be zero in this nominal condition due to the cylindrical symmetry of the system.},
\begin{equation}
F_z=m_z\,\,\partial B_z/\partial z\label{eq:Fz}.
\end{equation}
Note that the force can be also calculated as a reaction force, which is exerted on the current carried in the coil under an magnetic field made by the cylindrical magnet.

The on-axis field of a solenoid coil can be calculated by integrating all the contributions from every circular current loop in the coil, $B_z = \mu_0 Ia^2(a^2+z^2)^{-3/2}/2$, where $\mu_0$ is the magnetic permeability in vacuum, $a$ is a radius of each loop, and $I$ is the current carried in the wire. The location $z$, where $B_z$ is evaluated, is measured from the center point of the loop. The on-axis field $B_z$ of a single-layer $N$-turn solenoid having a radius of $a$ and a length of $L$ can be found in some textbooks as~\cite{Goto:1970,Jackson:1998}
\begin{equation}
B_z =\mu_0\frac{IN}{2L}\left(C(z+\tfrac{L}{2},a)-C(z-\tfrac{L}{2},a)\right),\label{eq:single_solenoid}\\
\end{equation}
where $C(x,a) \equiv x/(x^2+a^2)^{1/2}$, and the origin of $z$ is at the center point of the coil.

In the same manner, the on-axis field $B_z$ of a multilayer solenoid can be calculated as~\cite{Goto:1970,Duffy:1984}
\begin{widetext}
\begin{equation}
B_z = \mu_0\frac{NI}{2L(a_2-a_1)}\left(
\left(z+\frac{L}{2}\right)\ln{\left[\frac{a_2+\sqrt{a_2^2+(z+\frac{L}{2})^2}}{a_1+\sqrt{a_1^2+(z+\frac{L}{2})^2}}\right]}-
\left(z-\frac{L}{2}\right)\ln{\left[\frac{a_2+\sqrt{a_2^2+(z-\frac{L}{2})^2}}{a_1+\sqrt{a_1^2+(z-\frac{L}{2})^2}}\right]}
\right),\label{eq:multi_solenoid}
\end{equation}
\end{widetext}
where $a_1$ and $a_2$ are the inner and outer radii of the coil, respectively. Note that $N$ still represents the total number of turns of the wire; we consider a tiny area element $dr\,dz$ in the winding-wire region, where the number density of the wire turns was
$dN = N dr\,dz/(L(a_2-a_1))$, and so the corresponding current was $I\,dN$ in the area element.

Deriving the field gradient $\partial B_z/\partial z$ is trivial, but the resultant analytic form is complicated, so we do not show it here; recent computers can draw a graph of the analytic form quickly.
Note that $B_z$ becomes almost flat within the coil ($|z|<L/2$), and so $\partial B_z/\partial z \sim 0$. In addition, $\partial B_z/\partial z$ has two peaks of opposite sign around ends of the coil, $z\sim \pm L/2$; because $B_z$ is even, the derivative is odd. We could obtain $z$-locations of these peaks by solving $\partial^2 B_z/\partial z^2=0$, but that is difficult in analytic form.

A cylindrical magnet having a radius of $R$ and a height of $h$ provides an on-axis field $B_z$ at a location $z$ away from the center point of the magnet as~\cite{Furlani:2001}
\begin{equation}
B_z = \tfrac{1}{2}\mu_0M_z\left(C(z+\tfrac{h}{2},R)-C(z-\tfrac{h}{2},R)\right)\label{eq:cyl_mag},
\end{equation}
where $M_z$ is the magnitude of a uniform magnetization $\vec{M} = M_z\bm{e}_z$ of the magnet. The net magnetic flux density near the magnet is $\vec{B}=\mu_0(\vec{H}+\vec{M})$, where $\vec{H}$ is an external magnetic field, if any. Let $\vec{H}=0$ hereafter. Then the magnet provides the residual magnetic flux density (remanence): $\vec{B}_\mathrm{r}=\mu_0 \vec{M}$. 
We can measure the magnetic flux density of the surface on one end of the cylindrical magnet, $B_\mathrm{s}$, with a teslameter. Equating the measured $B_\mathrm{s}$ to Eq.~(\ref{eq:cyl_mag}) at $z=h/2$, we can obtain an estimate of $B_\mathrm{r}$:
\begin{equation}
B_\mathrm{r}=\mu_0 M_z = 2B_\mathrm{s}(1+(R/h)^2)^{1/2}.
\end{equation}
One can compare the estimate with a specification for the consistency provided by the magnet company. Finally, using the definition $M_z = m_z/V$, where $V=\pi R^2h$ is the volume of the magnet, we can obtain an estimate of $m_z$ as
\begin{equation}
m_z = 2B_\mathrm{s}\pi R^2(h^2+R^2)^{1/2}/\mu_0\label{eq:mz_mag}.
\end{equation}

The force in Eq.~(\ref{eq:Fz}) can be calculated by combining $\partial B_z/\partial z$ from Eq.~(\ref{eq:multi_solenoid}) and the estimate of $m_z$ in Eq.~(\ref{eq:mz_mag}).

Note that this estimation method uses Eq.~(\ref{eq:cyl_mag}) to involve the measured $B_\mathrm{s}$. If we use the field formula for a magnetic dipole at the origin~\cite{Jackson:1998}, or Eq.~(\ref{eq:Bz_dipole}), instead of Eq.~(\ref{eq:cyl_mag}), then, instead of Eq.~(\ref{eq:mz_mag}), the estimate would be $m_z = 2B_\mathrm{s}\pi(h/2)^3/\mu_0$ by substituting $z=h/2$ and $r=0$ into the dipole field; this loses the information of the magnet radius $R$.
Eq.~(\ref{eq:cyl_mag}) reduces to the dipole field only when $|z|\gg h$ and $|z|\gg R$. In fact, the estimates with Eq.~(\ref{eq:mz_mag}) matched better with the measured forces in our case.

\section{Eddy current loss in a coil bobbin}
\label{Sec:eddy_current_loss}
In this section, we will review how to estimate a viscous damping coefficient due to eddy current loss in the coil bobbin of a solenoid actuator. A similar discussion can be found in several documents~\cite{Cagnoli:1998,Barton:2000,Plissi:2004,Strain:2005}, but we want to summarize them into a short article.

Suppose a magnet and an electric conductor move slowly relative to each other.
A magnetic flux $\Phi$ from the magnet induces electromotive force $\mathcal{E}$ in the conductor: $\mathcal{E}=-d\Phi/dt$. Then let us consider magnetic flux across an area $S$ in the conductor, $\Phi = \int_S \vec{B}\cdot\vec{n}dS$, where $\vec{B}$ is the magnetic flux density, and $\vec{n}$ is the unit normal vector at the surface on $S$. The electromotive force induced in $S$ is written as~\cite{Jackson:1998}
\begin{equation}
\mathcal{E}=-\frac{d}{dt}\int_S\vec{B}\cdot\vec{n}dS = -\int_S\frac{\partial \vec{B}}{\partial t}\cdot\vec{n}dS
-\oint_C(\vec{B}\times\vec{v})\cdot d\vec{l},\label{eq:emf}
\end{equation}
where $C$ represents a contour of $S$, $d\vec{l}$ is a line element along $C$, and $\vec{v}$ is the velocity of $C$. Hereafter we will look at the system from the rest frame of the magnet, and so the first term of the right-hand side of Eq.(\ref{eq:emf}) becomes 0; only the second term, the contour integral along $C$, is left.

Let us consider the system consisting of the cylindrical magnet and the coil bobbin of the solenoid actuator and take the cylindrical coordinate in the same manner as in Appendix~\ref{Sec:force_by_coil}.
The coil bobbin moves in the $z$ direction, and so $\vec{v}=v_z\bm{e}_z$.
The line element can be written as $d\vec{l}=dl\,\bm{e}_\theta$.
Assuming $\vec{B}=B_r\,\bm{e}_r+B_\theta\,\bm{e}_\theta+B_z\,\bm{e}_z$ comes from a magnetic dipole, we obtain $B_\theta =0$ and the other components as~\cite{Jackson:1998}
\begin{align}
B_r&=\mu_0\frac{3m_z}{4\pi}\frac{rz}{(r^2+z^2)^{5/2}},\label{eq:Br_dipole}\\
B_z&=\mu_0\frac{m_z}{4\pi}\frac{2z^2-r^2}{(r^2+z^2)^{5/2}}\label{eq:Bz_dipole},
\end{align}
where the notation is the same as in Appendix~\ref{Sec:force_by_coil}; the origin of the coordinate is at the center point of the magnetic dipole.
Substituting these into Eq.~(\ref{eq:emf}) and noting that $\bm{e}_r\times\bm{e}_z =-\bm{e}_\theta$, the electromotive force is $\mathcal{E} = 2\pi r B_rv_z$, where  $r$ is a radius of the circular path for the contour integral. The electromotive force arises when a circular eddy current is induced around the central hole in the coil bobbin.

Consider a thin toroidal structure with a rectangular cross section in the coil bobbin at a radius of $r$ from the $z$-axis; the cross section is $dr\, dz$ and the circumference of the center of the toroid is $2\pi r$. The resistance of the toroid is $dR=\rho 2\pi r/(dr\, dz)$, where $\rho$ is the volume resistivity of the material of the bobbin. The Joule heating by the eddy current carried in the toroid is $dP = \mathcal{E}^2/dR$. Integrating $dP$ over the effective region of $r$ and $z$, the net energy loss by the eddy current is $P = 9\mu_0^2m_z^2 v_z^2 D/(8\pi\rho)$, where
\begin{equation}
D=\int_{r_1}^{r_2}\!\!\!\!\int_{z_1}^{z_2}\frac{r^3z^2}{(r^2+z^2)^5}dz\,dr.
\end{equation}
Here $z_1<z_2$ and $0<r_1<r_2$. Note that the integrand goes to zero rapidly if $r,z\rightarrow +\infty$; the lower limits $r_1$ and $z_1$ would mostly determine the value of $D$. Deriving an analytic form of $D$ is trivial~\footnote{Unlike ref.~\cite{Plissi:2004}, we limited our discussion to a system with cylindrical symmetry. By the way, ref.~\cite{Plissi:2004} missed substituting some parameters to its Eq.(A7).}, but the form is complicated, so we do not show it here. Today, a numerical integral is sufficiently fast and easily coded using suitable libraries such as \verb|scipy|, and coding the numerical integral is likely less error-prone than coding the complicated analytic form. In practice, the coil bobbin and the surrounding structures are not exactly cylindrical, so the calculation shown here is merely an approximation.

From Eq.~(\ref{eqs:two-body_mech}), the total mechanical energy of the two-body system is $E=\frac{1}{2}m_1 {\dot{x}_1}^2+\frac{1}{2}m_2 {\dot{x}_2}^2+\frac{1}{2}k_1x_1^2+\frac{1}{2}k_2x_2^2$. Then, the rate of change of the total energy with respect to time is calculated as  
$\frac{dE}{dt}=-c(\dot{x}_1-\dot{x}_2)^2$.
Equating the dissipated power $-dE/dt$ corresponding to $P$, we obtain a viscous damping coefficient $c=P/v_z^2$, or
\begin{equation}
c= \frac{9\mu_0^2m_z^2}{8\pi\rho}D \label{eq:viscous_eddy}
\end{equation}
due to the eddy current loss. If needed, one would include a term of viscous damping force with this coefficient into the equations of motion under consideration~\cite{Tsubono:1993}. 

\bibliography{KAGRA_OSEM}

\end{document}